\documentclass[prd,aps,twocolumn,nofootinbib,showpacs,superscriptaddress]{revtex4-1}
\usepackage{amsfonts}
\usepackage{amsmath}
\usepackage{amssymb}
\usepackage{bm}
\usepackage{dcolumn}
\usepackage{graphicx}
\usepackage{graphics}
\usepackage[latin1]{inputenc}
\usepackage{latexsym}
\usepackage{rotating}
\usepackage{hyperref}
\usepackage{xspace} % Sensible space treatment at end of simple macros
\usepackage[usenames]{color}
\usepackage{mathrsfs}
\usepackage{multirow}

\usepackage{longtable}
\usepackage{pict2e}
\usepackage[usenames,dvipsnames]{xcolor}
\usepackage{url}

\allowdisplaybreaks

\definecolor {darkgreen}{rgb}{0.2,0.7,0.2}
\definecolor{purple}{rgb}{0.5,0,0.5}

\newcommand{\apjl}{ApJ Letters} 

\newcommand{\bea}{\begin{eqnarray}}
\newcommand{\eea}{\end{eqnarray}}
\newcommand{\beq}{\begin{equation}}
\newcommand{\eeq}{\end{equation}}

\newcommand{\harm}{{\sc Harm3d}\xspace}

\def\fun#1#2{\lower3.6pt\vbox{\baselineskip0pt\lineskip.9pt
  \ialign{$\mathsurround=0pt#1\hfil##\hfil$\crcr#2\crcr\sim\crcr}}}

\def\lambdabar{%
\relax
\bgroup
\def\@tempa{\hbox{\raise.73\ht0
\hbox to0pt{\kern.25\wd0\vrule width.5\wd0
height.1pt depth.1pt\hss}\box0}}%
\mathchoice{\setbox0\hbox{$\displaystyle\lambda$}\@tempa}%
{\setbox0\hbox{$\textstyle\lambda$}\@tempa}%
{\setbox0\hbox{$\scriptstyle\lambda$}\@tempa}%
{\setbox0\hbox{$\scriptscriptstyle\lambda$}\@tempa}%
\egroup
}

\begin{document}

\title{Inspiralling, Non-Precessing, Spinning Black Hole Binary Spacetime
\\
via Asymptotic Matching}

\author{Brennan~Ireland}
\affiliation{Center for Computational Relativity and Gravitation, 
School of Mathematical Sciences, and School of Physics and Astronomy, Rochester Institute of
Technology, Rochester, New York 14623, USA.}

\author{Bruno~C.~Mundim}
\affiliation{Institut f\"ur Theoretische Physik,
Johann Wolfgang Goethe-Universit\"at, Max-von-Laue-Str. 1,
60438 Frankfurt am Main, Germany}

\author{Hiroyuki~Nakano}
\affiliation{Center for Computational Relativity and Gravitation, 
School of Mathematical Sciences, and School of Physics and Astronomy, Rochester Institute of
Technology, Rochester, New York 14623, USA.}
\affiliation{Department of Physics, Kyoto University,
Kyoto 606-8502, Japan.}

\author{Manuela~Campanelli}
\affiliation{Center for Computational Relativity and Gravitation, 
School of Mathematical Sciences, and School of Physics and Astronomy, Rochester Institute of
Technology, Rochester, New York 14623, USA.}

%%%%%%%%%%%%%%%%%%%%%%%%%%%%%%%%%%%%%%%%
\begin{abstract}

We construct a new global, fully analytic, approximate spacetime which
accurately describes the dynamics of non-precessing, spinning black hole
binaries during the inspiral phase of the relativistic merger process. This
approximate solution of the vacuum Einstein's equations can be obtained by
asymptotically matching perturbed Kerr solutions near the two black holes to a
post-Newtonian metric valid far from the two black holes. This metric is then
matched to a post-Minkowskian metric even farther out in the wave zone. 
The procedure of asymptotic matching is generalized
to be valid on all spatial hypersurfaces, instead of a small group
of initial hypersurfaces discussed in previous works. 
This metric is well suited for long term dynamical simulations of
spinning black hole binary spacetimes prior to merger, such as studies of
circumbinary gas accretion which requires hundreds of binary orbits.

\end{abstract}

\maketitle

%%%%%%%%%%%%%%%%%%%%%%%%%%%%%%%%%%%%%%%%
\section{INTRODUCTION}

%%%%%%%%%%%%%%%%%%%%%%%%%%%%%%%%%%%%%%%%

The recent announcement of GW150914~\cite{GWdetect} from the 
Laser Interferometer Gravitational wave Observatory (LIGO)~\cite{2009RPPh...72g6901A}
has provided the first strong evidence of a black hole binary (BHB) coalescence due to the
emission of gravitational waves (GWs).  This discovery kicks off the era of gravitational 
wave astronomy, and provides further justification to the study of the inspiral and merger of BHBs.
The merger of stellar-mass black holes (BHs), such as GW150914 and other BHB potential LIGO sources, 
are not expected to have any electromagnetic (EM) counterparts~\cite{2016ApJ...818L..22A}.
However,  Ref.~\cite{CONN16} reported a 
potential $\gamma$-ray EM counterpart observed by the Gamma-Ray Burst Monitor~\cite{GBM} on Fermi which is
consistent with the sky localization of GW150914.

The existence of supermassive BHs residing in the centers of
galaxies~\cite{2009ApJ...698..198G} indicates another type of possible BHB, one for which
EM counterparts are not only possible, but are also frequent. 
Galaxies will undergo 
mergers of both stars and gas as the Universe evolves~\cite{1985Natur.317..595F, 1986ApJ...304...15B}.
In the process of the galactic merger, the supermassive BHs will form a bound pair due to torques 
exerted by the surrounding gas and stars, dynamical friction, and gravitational slingshots that eject stars 
from the nucleus of the merging system.
Gravitational radiation becomes the dominant energy loss mechanism as the BHBs become
closely separated, eventually driving the two BHs to the full merger. 
The consequences for galactic evolution of these mergers 
are very deep, as strong correlations
between galactic structure and central BH mass indicate tight feedback
between BH and galaxy growth.

Future space-based GW missions,
like the proposed European New Gravitational wave Observatory
(NGO)~\cite{2013GWN.....6....4A,2012CQGra..29l4016A,2013arXiv1305.5720C}
and the DECi-hertz Interferometer Gravitational wave Observatory
(DECIGO)~\cite{Seto:2001qf}, will be sensitive to such events, but are decades away from launch.
Fortunately, in the case of supermassive BHBs, 
highly-relativistic magnetized gas could flow around the pair, as
well as around each BH companion. Therefore, powerful
EM signals should accompany the inspiral and merger 
of BHBs~\cite{Roedig:2014cea,2013CQGra..30x4007S}.

The main research focus of the authors is to simulate the effects of spinning
supermassive BHB mergers on nearby gas in sufficient detail to enable EM 
observations of these events.
This paper represents a major step forward in achieving this goal by providing an analytic 
spacetime that can handle spinning BHBs.
Since 2005~\cite{Pretorius:2005gq, Campanelli:2005dd, Baker:2005vv}, a handful
of numerical relativity simulations of BHBs have been successfully carried on for nearly a hundred
orbits~\cite{Lousto:2015PRL,Szilagyi:2015rwa}, providing the necessary 
waveforms for current and future GW
detectors~\cite{Aasi:2014tra,Mroue:2013xna}. 

However, in the case of BHBs in a gaseous environment, 
numerical magnetohydrodynamic (MHD) simulations are still very expensive 
to carry out~\cite{Bode:2010ApJ,Bode:2012ApJ,Pal:2010MNRAS,Farris:2010PhRvD,
Farris:2011PhRvD, Farris:2012PhRvL,Farris:2014ApJ,
Farris:2015MNRAS,Giacomazzo:2012ApJ,Gold:2013APS,Gold:2014PhRvD}.
This is because we 
need to resolve turbulences and shocks in the gas, as well as secular variations 
in the circumbinary disk on the
time scale of hundreds to thousands of binary orbits (see Ref.~\cite{Noble:2012xz} for detailed 
discussion). In order to make long-term and accurate MHD simulations possible, we developed a
complementary analytic approach to treat dynamical, nonspinning, BHB
spacetimes~\cite{Noble:2012xz, Mundim:2013vca,
Zilhao:2014ida, Zilhao:2013dta}. This spacetime is a solution to 
the Einstein field equations in the approximation that the BHB is slowly inspiralling 
to merger. 
In this situation, gravity is weak
[$r_g/r = GM/(rc^2) \ll 1$] and motions are slow [$v/c \ll 1$], so the
post-Newtonian (PN) approximation is a very good description of spacetime. 
Using a spacetime accurate to 2.5PN order
(i.e., including terms up to $\sim (r_g/r)^{5/2}$), and using the 3.5PN
equations of motion (EOM) to describe the GW driven inspiral of the 
orbital evolution~\cite{Blanchet:2013haa}, 
we demonstrated that circumbinary disks can track
a supermassive BHB for hundreds of orbits until the binary practically reaches the relativistic
merger regime~\cite{Noble:2012xz}. 

In a more recent paper~\cite{Mundim:2013vca}, we extended the metric all the way down 
to the horizons of each BH. We did this by broadening
the framework introduced in Refs.~\cite{Yunes:2005nn, Yunes:2006iw,
JohnsonMcDaniel:2009dq} for constructing a spacetime metric valid for initial
data, to a full dynamical spacetime metric valid for arbitrary times.
This metric is constructed by stitching together different spacetime metrics
valid in different regions of the full BHB spacetime (see~\ref{sec:global-metric}).
We extend this framework here to include the effects of spinning BHs.

There are important spin-based effects which 
affect the dynamics of the BHB that can also significantly alter the dynamic of the surrounding gas.
The mechanisms associated with accretion at larger separations may
drive the spins into alignment with both the binary orbital axis and the
circumbinary disk axis~\cite{Bogdanovic:2007ApJ,Miller:2013ApJ,Sorathia:2013ApJ}.  
In this case, another mechanism due to spin-orbit coupling can delay or
prompts the merger of the BHB according to the sign of the spin-orbit
coupling~\cite{Campanelli:2006uy}.
This could have a significant effect on the total pre-merger light output and its time-dependence. 

At closer BHB separations, little is know about the effectiveness of these accretion driven spin-alignment mechanisms.
In this situation, spin-spin and spin-orbit
interactions can also cause the spins to precess, leading to time-dependence in
non-planar gas orbits~\cite{Campanelli:2006PhRvD}. Gravitomagnetic torques 
arising from the BH spins oblique to the orbital axis may then push the accretion 
streams onto the BHs out of the orbital plane and alter the
tidal limitation of the mini-disks, particularly for relatively small binary
separations. Another interesting spin effect is the recently discovered spin-flip-flop 
phenomenon~\cite{Lousto:2015PRL,Lousto:2015uwa}, where the spin of one of the BHs completely reverses. 
This may cause
the gas in the neighborhood of the binary to be continually disturbed in a manner 
that will produce a very distinct EM signature from the disk. Finally,
highly spinning BHBs may recoil at thousands of
km/s~\cite{Campanelli:2007ew,Campanelli:2007apj,Koppitz:2007prl,Gonzalez:2007prl} due to asymmetrical emission of gravitational radiation induced by the BH spins~\cite{Lousto:2012PhRvD,Lousto:2011kp}. 
The resulting ejected BH may carry along part of the original accretion 
disk causing it to be bright enough to be observable (see~\cite{Komossa:2012cy} for a review).

In this paper, we generalize~\cite{Mundim:2013vca}
to spinning BHBs, with spins aligned
and counteraligned with the orbital angular momentum of the binary, in a quasi
circular inspiral. We will address the case of oblique spins and spin
precession~\cite{Thorne:1985PRD} in future work~\cite{Nakanoetal_in_prep}. 
Our new spinning global metric must, of course, approximately satisfy the Einstein
equations, if it is to be considered a true spacetime metric representing a
BHB. For each zone, we check the validity of the spacetime
analytically in the black hole perturbation, the PN
and the post-Minkowski (PM) approximations by computing the deviations from
Einstein's equations.
We can construct several curvature
invariants to determine the overall accuracy of the
approximations. One such invariant is the Ricci scalar, which can be
compared against the exact vacuum solution quantity of $R=0$. Another
quantity is the Hamiltonian constraint, which is used in the numerical
relativity community to measure the amount of ``fake" mass in the system caused
by violations to the Einstein vacuum field equations. Finally, we introduce an
invariant quantity related to the Kretschmann invariant $R_{\mu \nu \rho
\delta} R^{\mu \nu \rho \delta}$, which has the benefit of being a normalized
measure of the violation of the global metric to the Einstein equations.

This paper is organized as follows. Sec.~\ref{sec:global-metric} outlines the
different approximate metrics, and details of their construction and matching
to obtain the global metric. Sec.~\ref{sec:numerical} discusses the numerical
analysis of this global metric by the calculation of several spacetime
invariants that impress upon us the validity of the global metric. Finally,
Sec.~\ref{sec:discussion} contains useful discussion, conclusions, and future
work. The appendices~\ref{app:trans_func}, \ref{app:IK_to_CSH}, \ref{app:ISCO}
and \ref{app:IZtreatment} describe the choice of transition functions that are
utilized in the global metric, the details of the ingoing Kerr to Cook-Scheel
coordinate transformation, the innermost stable circular orbit (ISCO) and an effective
evaluation of the inner zone metric.

Throughout this paper, we follow the notation of Misner, Thorne and
Wheeler~\cite{MTW}, specifically, greek letters ($\alpha, \beta, \gamma, \cdot\cdot\cdot$) used as
indices are indicative of spacetime coordinates, and latin letters ($i,j,k, \cdot\cdot\cdot$) 
are used in discussions of spatial coordinates only. The covariant metric
is then written as $g_{\mu \nu}$, and has a signature of $(-,+,+,+)$.  We use
the geometric unit system, where $G=c=1$, with the useful conversion factor 
$1 M_{\odot} = 1.477 \; {\rm{km}} = 4.926 \times 10^{-6} \; {\rm{s}}$.

%%%%%%%%%%%%%%%%%%%%%%%%%%%%%%%%%%%%%%%%
\section{CONSTRUCTION OF APPROXIMATE GLOBAL METRIC}
\label{sec:global-metric}
%%%%%%%%%%%%%%%%%%%%%%%%%%%%%%%%%%%%%%%%

We are concerned with the construction of the approximate global metric with
spin for a BHB pair on a quasi circular inspiral, in the inspiral regime. To
find this global metric for the BHBs, we first consider the individual regions
where different approximations and assumptions hold (see Table~\ref{tab:zones});
the inner zone (IZ) around BH1 (IZ1) and around BH2 (IZ2),
the near zone (NZ) around the two BHs, and
the far zone (FZ) or wave zone farthest out.

\begin{table}[!ht]
\caption{Regions of validity for the different zones and BZ locations.
Here $r_1$ and $r_2$ are the distances from the first or second BH with
mass $m_1$ or $m_2$, $r$ is the distance from the center of mass to a field
point, $r_{12}$ is the orbital separation, and $\lambda$ is the gravitational
wavelength. For BZs to exist, the system must satisfy $m_{1,2} \ll r_{12}$,
though we expect that the metric will break down before this condition is
violated. (This was also presented in Refs.~\cite{Gallouin:2012kb,
Mundim:2013vca,Zlochower:2015baa}.)}
\label{tab:zones}
\begin{center}
\begin{tabular}{c|ccccc}
  \hline\hline
  Zone     & \multicolumn{5}{c}{Region of Validity}\\
  \hline
  IZ1      & $0     $&$  <    $&$  r_1  $&$  \ll  $&$ r_{12}$  \\
  IZ2      & $0     $&$  <    $&$  r_2  $&$  \ll  $&$ r_{12}$ \\
  NZ       & $m_A   $&$  \ll  $&$  r_A  $&$  \ll  $&$ \lambda$ \\
  FZ       & $r_{12} $&$  \ll  $&$  r    $&$  <    $&$ \infty$ \\
  IZ1-NZ BZ & $m_1   $&$  \ll  $&$  r_1  $&$  \ll  $&$ r_{12}$  \\
  IZ2-NZ BZ & $m_2   $&$  \ll  $&$  r_2  $&$  \ll  $&$ r_{12}$  \\
  NZ-FZ BZ & $r_{12} $&$  \ll  $&$  r    $&$  \ll  $&$ \lambda$ \\
  \hline\hline  
\end{tabular}
\end{center}
\end{table}

%%%%%%%%%%%%%%%%%%%%
\subsection{Subdividing spacetime}
%%%%%%%%%%%%%%%%%%%%

In the inner region very close to the individual BHs, 
we treat the spacetime as a vacuum Kerr solution with
linear perturbations as in Ref.~\cite{Yunes:2005ve}. The PN metric subdivision
that was briefly discussed above is known as the NZ. This
metric is valid in the slow motion, weak field limit. The addition of spins to
the NZ will add terms to the PN expansion which, to lowest order, are the
1.5PN~\footnote{A PN order $N$ is said to be a term of order $(v/c)^{2N}$ for
the slow motion expansion (e.g., 1.5PN is order $(v/c)^3$).} leading order
spin-orbit coupling and 2PN leading order spin-spin terms. The NZ is defined
to be valid in a region far away from the individual BHs (to not violate the
weak field approximation), but not farther away than a gravitational wavelength
($\lambda \sim 2 \pi/\omega_{\rm GW} \sim \pi / \omega_{\rm orb} \sim \pi(r_{12}^3/M)^{1/2}$, 
where $M = m_1 + m_2$ is the total mass) from
the center of mass of the binary system. 
The region even farther out than a gravitational wavelength
from the center of mass can be described as the FZ, 
in which the metric takes the form of flat (Minkowski) space, with
outgoing GWs perturbing the spacetime. The FZ is modeled with a
PM (or multipolar) formalism~\cite{Will:1996zj}. Unlike in PN
formalism, PM expansions correctly treat the retardation of the gravitational
field, which is essential for understanding the FZ. This subdivision of the
spacetime into different regions will only be valid as long as the slow motion
approximation holds, and will break down around an orbital separation $r_{12}
\approx 10M$~\cite{Yunes:2008tw,Zhang:2011vha}. 

Once we have the individual metrics for the different zones, we need to stitch
them all together into a global metric, asymptotically matching the IZ, NZ, and
FZ to each other in BZs. The procedure of matching adjacent metrics to one
another requires the metrics to be in the same coordinate system. In other
words, one of the metrics (and its parameters) will be related to the next door
metric via some coordinate transformations, constructed such that the
transformed metric asymptotes to the adjacent metric in the BZ. Asymptotic
matching in GR has been successfully done in Refs.~\cite{Yunes:2005nn,
Alvi:1999cw, Pati:2002ux, Alvi:2003pn, Yunes:2006iw, Yunes:2006mx,
JohnsonMcDaniel:2009dq, Gallouin:2012kb}, but in all of these papers, the authors asymptotically
matched in the context of initial data for BHB simulations, which implies that
their focuses were on a particular spatial hypersurface. However, in the
context of this work, this restriction must be lifted if there is to be any
hope of dynamic, long time evolutions of BHBs. Ref.~\cite{Mundim:2013vca}
successfully removed this restriction in the context of nonspinning BHs. The
task now is to do this in the context of spinning BHs.  The extension to
spinning BHs should be more astrophysically relevant than the non spinning BHB
case covered in Ref.~\cite{Mundim:2013vca}, because it is thought that most
astrophysical BHs have spin~\cite{Bardeen:1970Natur}.

Once the metrics have been asymptotically matched, we construct a global metric
by introducing transition functions that take us from one
metric to the next in the BZs without introducing artificial
errors~\cite{Yunes:2006mx} into the metric that are larger than the errors
already incurred in the approximations used in construction of the individual
metrics.

\begin{figure}[!ht]
\begin{center}
\includegraphics[width=\columnwidth,clip=true]{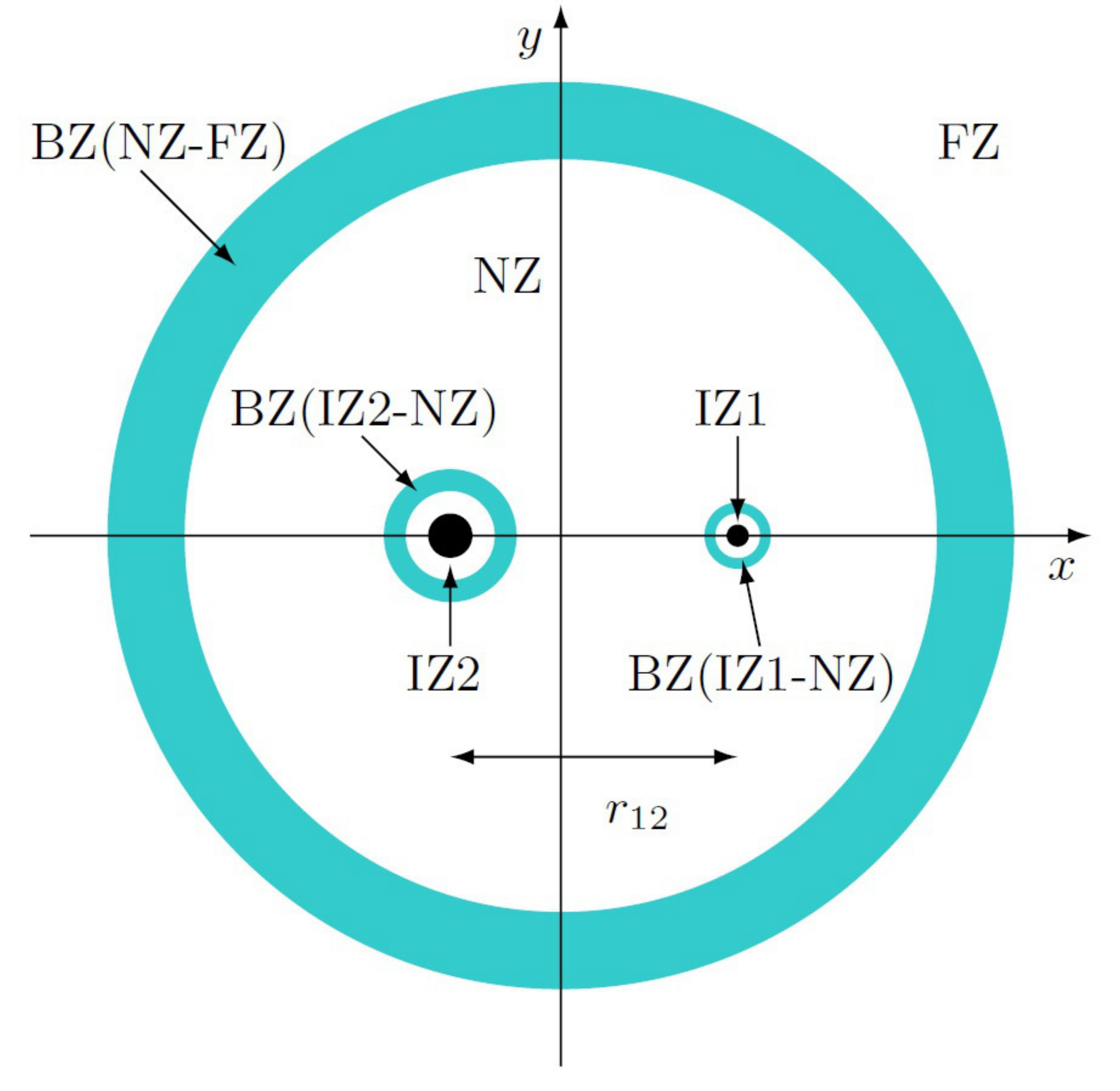}
\end{center}
\caption{A schematic diagram detailing the different zones 
for the approximate analytic spacetime looking down the $z$-axis 
at a particular instant in time. The black dots 
on the $x$-axis represent the two BHs, 
with an orbital separation of $r_{12}$. 
The cyan shells indicate the BZs 
-- regions where two adjacent metrics have overlapping regions of validity. 
The outermost shell is the NZ-FZ BZ, 
with both the IZ1-NZ and IZ2-NZ BZs labeled 
around the individual BHs. The IZ, NZ, and FZ are denoted 
as the regions contained by the BZs. 
Note that the circular nature of the BZs is not physical, 
only schematic, and in general it is expected 
that they will have some distortions.
(This was also presented in Refs.~\cite{Gallouin:2012kb, Mundim:2013vca,Zlochower:2015baa}.)}
\label{fig:zones}
\end{figure}

The different zones and their associated BZs are summarized in
Fig.~\ref{fig:zones} and Table~\ref{tab:zones}. The cyan shells indicate the
BZs between the individual subdivided metrics, where both of the adjacent
metrics are valid. We note that the figure is purely schematic; in general,
there is no inherent symmetry that would cause the BZs to be like the spherical
shells depicted, so in reality these BZs would be distorted.  It is in these
BZs that asymptotic matching of  the individual metrics takes place.  Finally,
the resulting matched metrics are stitched together with the proper transition
functions satisfying the Frankenstein theorems~\cite{Yunes:2006mx}, yielding a
global analytic approximate spacetime. The construction of the asymptotically
matched global metric has been calculated in Ref.~\cite{Gallouin:2012kb}, but
only in the context of initial data and not for long time evolutions of the BHB
system.

%%%%%%%%%%%%%%%%%%%%
\subsection{Inner Zone}\label{sec:IZ}
%%%%%%%%%%%%%%%%%%%%

The IZs in Fig.~\ref{fig:zones} and Table~\ref{tab:zones}
are constructed following the work laid out in 
Ref.~\cite{Yunes:2005ve}, and applied to BHB initial data 
on a single spatial hypersurface in Ref.~\cite{Gallouin:2012kb}. 
The IZ metric is approximately described by the Kerr metric 
$g_{\mu \nu}^{\rm Kerr}$ plus a linearized vacuum perturbation 
$h_{\mu \nu}^{\rm IZ}$:
\begin{align}
g_{\mu \nu}^{\rm IZ} = g_{\mu \nu}^{\rm Kerr} + h_{\mu \nu}^{\rm IZ} \,.
\end{align}
Here the Kerr metric is given by the mass of the Kerr BH $m_{\rm Kerr}$,
and the dimensional spin parameter $a$, which can be related 
to the dimensionless spin parameter $\chi$ by $a = \chi m_{\rm Kerr}$ 
and the dimensional spin $S = \chi m_{\rm Kerr}^2$. 
It is convenient to work with $\chi$ in our calculations because 
it is a normalized quantity ($0 \leq |\chi| \leq 1$), 
with zero being a non spinning (Schwarzschild) BH, 
and one being a maximally spinning Kerr BH. 

The metric perturbation $h_{\mu \nu}^{\rm IZ}$ has been studied 
and applied for Schwarzschild BHs, where we can use 
the Regge-Wheeler-Zerilli-Moncrief 
formalism~\cite{Regge:1957td, Zerilli:1971wd, Moncrief:1974am,Nagar:review}, 
which is generally valid for static spherically symmetric spacetimes. 
However, when looking at the case of the Kerr background, 
this formalism is not applicable, because Kerr is not spherically symmetric, 
and so there is not a multipole decomposition of metric perturbations, 
and the Einstein equations cannot be uncoupled into wave equations. 
A reformulation due to Newman and Penrose~\cite{Newman:1961qr} 
(from here out referred to as the NP formalism) of the Einstein equations 
and Bianchi identities projected along a null tetrad coinciding 
with the null symmetries of the spacetime 
(Kerr BHs are a type D algebraically special solution) allowed 
Teukolsky~\cite{Teukolsky:1973ha} to write down
a single master wave equation for the perturbations of Kerr 
in terms of the Weyl scalars (constructed from contracting 
the Weyl tensor with the same conveniently chosen null tetrad, 
called the Kinnersley tetrad) $\psi_0$ or $\psi_4$. Solutions 
to the Teukolsky equation yield the perturbed Weyl scalars.

To obtain the metric perturbation from the Weyl scalar, 
we must use the Chrzanowski procedure~\cite{Chrzanowski:1975wv}, 
which takes the Weyl scalar $\psi_0$ and acts a differential operator 
on it to yield a valid metric perturbation. 
This was later amended by Wald~\cite{Wald:1978vm}, and Kegles and 
Cohen~\cite{Kegeles:1979an} to use a Hertz potential $\Psi$ 
instead of a Weyl scalar ($\psi_0$ for ingoing radiation, 
suitable for studying perturbations, or $\psi_4$ for outgoing radiation,
suitable for GW studies~\cite{Campanelli:1998jv}).
A brief description 
of the metric perturbation construction via the Chrzanowski procedure 
is summarized in the following.

The metric perturbation $h_{\mu \nu}$ is constructed via the Chrzanowski
procedure by applying a certain differential operator to the so-called Hertz potential.
The Hertz potential must satisfy a certain differential equation 
with a source given by the NP scalar $\psi_0$,
and the differential equation can be inverted to yield the potential 
$\Psi$ totally in terms of $\psi_0$~\cite{Wald:1978vm, Ori:2002uv}. 
Therefore, the construction of the metric perturbation $h_{\mu \nu}$ 
boils down to finding an appropriate solution for the NP scalar $\psi_0$.
The metric perturbation is in the ingoing radiation gauge, given by
the perturbation contracted along the tetrad components,
$h_{\ell \ell} = h_{\ell n} = h_{\ell m} =  h_{\ell \overline{m}} = h_{m \overline{m}} = 0$
where $\ell_{\mu}$ and $m_{\mu}$ are components of the Kinnersley null tetrad,
and $\overline{m}_{\mu}$ is the complex conjugate of  $m_{\mu}$.

This NP scalar must, of course, satisfy the Teukolsky 
equation~\cite{Teukolsky:1973ha}. However, when the external universe 
(the source of the perturbation) is slowly-varying 
(as is the case of most interest to this work, when the external universe 
is a second BH on a quasi-circular inspiral with a large separation), 
it is possible to solve this equation perturbatively~\cite{Yunes:2005ve}. 
Thus, we can write $\psi_0$ in terms of the spin-2 weighted spherical
harmonics $_{2}Y_{l m}$ as
\begin{align}
\psi_{0} = \sum_{\ell,m} R_{\ell m}(r) z_{\ell m}(v) \; {}_{2}Y_{\ell m}(\theta,\phi) \,,
\end{align}
where the radial and time dependence are product decomposed into terms of
unknown real functions $R_{\ell m}(r)$, and complex functions $z_{\ell m}(v)$.
Here, $v$ is the advanced Kerr-Schild time coordinate. 
$z_{\ell m}$ can be written in terms of electric and magnetic tidal tensors,
which to leading order in BH perturbation theory~\cite{Poisson:2004cw} 
can be truncated at the $\ell = 2$ quadrupolar deformation. 
The radial functions $R_{\ell m}$ must then satisfy the 
(time-independent) Teukolsky equation, and can be solved in terms of
hypergeometric functions. Solving the time-independent Teukolsky equation
implies that the $z_{\ell m}$ functions are slowly varying or constant
in time. This can then be reconstructed into a functional form for $\psi_0$. 

With this NP scalar under control, it is possible to compute 
the Hertz potential, and from that, the full metric perturbation 
$h_{\mu \nu}$. Ref.~\cite{Yunes:2005ve} provides the full form of 
$h_{\mu \nu}$ in Eddington-Finkelstein (ingoing Kerr) coordinates.

The final form of the IZ metric needs to have several desirable features 
to be of use. One of which is horizon penetrating, Cook-Scheel harmonic 
(CS-H) coordinates ($T,\,X,\,Y,\,Z$)~\cite{Cook:1997qc}. 
The details of taking the IZ from the IK coordinates to the more useful 
CS-H coordinates is left to Appendix~\ref{app:IK_to_CSH}. From the CS-H
coordinates, it is simply a matter of applying yet another transformation 
to take the metric from the IZ coordinates to the coordinates used 
in the NZ, the PN harmonic (PNH) coordinates.

%%%%%%%%%%%%%%%%%%%%
\subsection{Near Zone}\label{sec:NZ}
%%%%%%%%%%%%%%%%%%%%

The NZ in Fig.~\ref{fig:zones} and Table~\ref{tab:zones}
is the region sufficiently distant from either BH 
that the metric can be described through the PN metric:
\begin{align}
g_{\mu \nu}^{\rm NZ} = \eta_{\mu \nu} + h_{\mu \nu}^{\rm NZ} \,,
\end{align}
where $\eta_{\mu \nu}$ is the Minkowski (flat space) metric, 
and $h_{\mu \nu}$ is a PN metric perturbation~\cite{Gallouin:2012kb}. 
In the PN approximation, the Einstein equations are solved 
in an expansion of both $v/c \ll 1$ (slow motion) and 
$G M/(R c^2) \ll 1$ (weak fields), where $M$ is the total mass 
of the BHs, $R$ is the orbital separation ($r_{12}$) \emph{or} 
the center of mass to one BH ($r_{1,2}$), and the $G$'s and $c$'s 
have been replaced for convenience. By construction, 
the PN approximation models the BHs as point particles.

The metric perturbation we use is specified in Ref.~\cite{Blanchet:1998vx} 
for the spin independent terms and the first non-vanishing spin terms 
are outlined in Refs.~\cite{Tagoshi:2000zg, Faye:2006gx}, 
giving us a 1.5PN metric in PNH coordinates 
(which are a Cartesian-like, rectangular coordinate system) 
to match the IZ to as
\begin{align}
g_{00}^{\rm NZ} +1 =&
\frac{2 m_{1}}{r_{1}} + \frac{m_{1}}{r_{1}} 
\left[4 \mathbf{v}_{1}^{2} - (\mathbf{n}_{1} \cdot \mathbf{v}_{1})^{2} \right]
- 2 \frac{m_{1}^{2}}{r_{1}^{2}} 
\cr &
- m_{1} m_{2} 
\Biggl[ \frac{2}{r_{1} r_{2}} + \frac{r_{1}}{2 r_{12}^{3}}
- \frac{r_{1}^{2}}{2 r_{2} r_{12}^{3}} + \frac{5}{2 r_{2} r_{12}} \Biggr]
\cr
& + \frac{4 m_{1} m_{2}}{3 r_{12}^{2}} (\mathbf{n}_{12} \cdot \mathbf{v}_{12}) 
\cr
& + \frac{4}{r_1^2} \epsilon_{ijk} v_1^i s_1^j n_1^k 
+ (1 \leftrightarrow 2) + {\cal{O}}(v^6)
\,,
\cr
g_{0i}^{\rm NZ} =&
-\frac{4 m_{1}}{r_{1}} v_{1}^{i} 
-\frac{2}{r_1^2} \epsilon_{ijk} s_1^j n_1^k
+ (1 \leftrightarrow 2) + {\cal{O}}(v^5)
\,,
\cr
g_{ij}^{\rm NZ} - \delta_{ij} =& \frac{2 m_{1}}{r_{1}} \delta_{ij}
+ (1 \leftrightarrow 2) + {\cal{O}}(v^4)
\,,
\label{eq:NZmetric}
\end{align}
where $m_{A}$, $s_A^i$, $y_{A}^{i}$ and $v_{A}^{i}$
denote the mass, spin angular momentum, location and velocity 
of the $A$th PN particle,
respectively.
Other notations that have been introduced above are
$r_{12} = |\mathbf{y}_{1}-\mathbf{y}_2|$, 
$\mathbf{n}_{12} = (\mathbf{y}_{1} - \mathbf{y}_{2})/b$, 
$\mathbf{v}_{12} = \mathbf{v}_{1} - \mathbf{v}_{2}$,
$r_A=|\mathbf{x}-\mathbf{y}_A|$, $n_A^i=(x^i - y_A^i)/r_A$
and $\epsilon_{ijk}$ is the Levi-Civita symbol.

In practice, we use higher order PN EOM 
than is strictly allowed by matching. Since the IZ metric is 
a first order BH perturbation theory, we cannot use any higher order 
than linear for the NZ metric in the matching calculation. 
We can, however, use a higher order PN EOM outside of the matching 
to have more accurate PN dynamics in the NZ for long time evolutions 
of BHB systems. The higher order formulas are summarized 
in the appendix of Ref.~\cite{Ajith:2012tt}, and also 
in Ref.~\cite{Blanchet:2013haa} (see also Sec.~\ref{sec:hangup}).
More specifically, we may use 
the energy function in Eq.~(A.11), flux function in Eq.~(A.13),
and mass loss in Eq.~(A.14) given in Ref.~\cite{Brown:2007jx}
and follow the procedure to derive the orbital phase evolution
presented in Sec.~9.3 of Ref.~\cite{Blanchet:2013haa}.
This gives the aligned spinning version of Eq.~(317) in Ref.~\cite{Blanchet:2013haa}
which is for the nonspinning case.

%%%%%%%%%%%%%%%%%%%%
\subsection{Asymptotic Matching}\label{sec:AM}
%%%%%%%%%%%%%%%%%%%%

The IZ metric is described by the CS-H coordinates $X^{\alpha}$ and the
parameters $\Lambda^{\alpha} = (m_{\rm Kerr},\,a,\,z_{R,m},\,z_{I,m})$, where $z_{R,m}$
and $z_{I,m}$ are the real and imaginary parts of $z_{2 m}$ respectively.  On
the other hand, the NZ metric is written in PNH coordinates $x^{\alpha}$ with the parameters $\lambda^{\alpha}
= (m_{1},\,m_{2},\, b,\, s_{1}^{i},\, s_{2}^{i})$.
We require that these two expressions be diffeomorphic to each other, leading us to a set of 
equations that relate the coordinates of the two metrics
\begin{align}
g^{\rm NZ}_{\alpha \beta} = \frac{\partial X^{\gamma}}{\partial x_{\alpha}} 
\frac{\partial X^{\delta}}{\partial x_{\beta}} 
g^{\rm IZ}_{\gamma \delta} \,,
\label{eq:coord_transf}
\end{align}
and expressions that relate the parameters used in each zone.
We consider $b=r_{12}$ a constant in the matching calculation here and
recover the time dependence in the final expression.

In the BZ in Fig.~\ref{fig:zones} and Table~\ref{tab:zones},
we use series expansions with respect to $(m_{2}/b)^{1/2}={\cal{O}}(v)$.
The IZ coordinates and parameters are expanded as
\begin{align}
X^{\alpha}\left( x^{\beta} \right) =& \sum_{i=0}^{n} \left( \frac{m_{2}}{b} \right)^{i/2}
 \left( X^{\alpha} \right)_{i} \left( x^{\beta} \right)
+ {\cal{O}}( v^{n+1} ) \,,
\nonumber \\
\Lambda^{\alpha}\left( \lambda^{\beta} \right) =& \sum_{i=0}^{n} \left( \frac{m_{2}}{b} \right)^{i/2}
 \left( \Lambda^{\alpha} \right)_{i} \left( \lambda^{\beta} \right)
+ {\cal{O}}( v^{n+1} ) \,,
\end{align}
where $(X^{\alpha})_{i}$ and $(\Lambda^{\alpha})_{i}$
denote $i$th expansion functions of the NZ coordinates $x^{\beta}$
and those of the NZ parameters $\lambda^{\beta}$, respectively.

In the asymptotic matching between the IZ1 and NZ metrics
to $O[(m_{2}/b)^1]$, i.e., $n=2$ in the above equations,
we have already discussed in Ref.~\cite{Gallouin:2012kb}
that the non-spinning matching transformation is sufficient,
even if we consider the matching of the spinning case.
This is because the spinning body effect arises from the $n=3$ matching. 
Obtaining the mass $m_{\rm Kerr}=m_{1}$ 
(also the dimensional spin parameter
$a=s_{1}^{z}/m_{1}$ for non-precessing, spinning BHBs),
the quadrupolar field is
\begin{align}
z_{R,0} &= \frac{2 m_2}{b^3} \,,
\cr
z_{R,2} &= \frac{6 m_2}{b^3} \cos 2\omega t \,, \quad
z_{R,-2} = \frac{6 m_2}{b^3} \cos 2\omega t  \,,
\cr
z_{I,2} &= - \frac{6 m_2}{b^3} \sin 2\omega t \,, \quad
z_{I,-2} = \frac{6 m_2}{b^3} \sin 2\omega t \,,
\end{align}
where $\omega \sim \sqrt{M/b^3}$ to lowest PN order,
and the other components vanish.

%%%%%%%%%%%%%%%%%%%%
\subsection{Expansion of the Nonspinning Part of the IZ and NZ Metrics}\label{sec:expIZNZ}
%%%%%%%%%%%%%%%%%%%%

Using $m_{1} \ll r_1 \ll b$ in the BZ in Fig.~\ref{fig:zones} and Table~\ref{tab:zones},
we expand the NZ and IZ metrics.
First, the NZ metric is expanded as in Ref.~\cite{JohnsonMcDaniel:2009dq}
\begin{align}
g_{\alpha \beta} =& (g_{\alpha \beta})_{0}
+ \sqrt{ \frac{m_{2}}{b} }(g_{\alpha \beta})_{1}
+ \left( \frac{m_{2}}{b} \right) (g_{\alpha \beta})_{2}
+ {\cal{O}}(v^3) \,,
\end{align}
where
\begin{align}
(g^{\rm NZ}_{\alpha \beta})_{0} =& \eta_{\alpha \beta} \,,
\quad
(g^{\rm NZ}_{\alpha \beta})_{1} = 0 \,,
\cr
(g^{\rm NZ}_{\alpha \beta})_{2} =& \biggl[ \frac{2m_{1}}{m_{2}} \frac{b}{(r_{1})_{0}}
+ 2 - \frac{2}{b} \left\{(\mathbf{r}_{1})_{0} \cdot (\mathbf{\hat b})_{0}\right\}
\cr &
+ \frac{1}{b^{2}} \left\{3 [(\mathbf{r}_{1})_{0} \cdot (\mathbf{\hat b})_{0}]^{2} - [(r_{1})_{0}]^{2}\right\}
\biggr]
\Delta_{\alpha \beta} \,.
\label{eq:NZ_bz}
\end{align}
Here, $\Delta_{\alpha \beta} = \mbox{diag}(1,\,1,\,1,\,1)$, and $(\hat b^k)_{0}=\hat
\beta^k=\{\cos\omega t,\,\sin\omega t,\,0 \}$ is a unit vector.  Note that
there is no spin contribution which is 1.5PN order.

Next, we treat the IZ metric in the BZ. The IZ metric
up to the second order is derived as
\begin{align}
(g^{\rm IZ}_{\alpha \beta})_{0} =& \eta_{\alpha \beta}
\, \quad
(g^{\rm IZ}_{\alpha \beta})_{1} = 0
\,, \cr
(g^{\rm IZ}_{00})_{2} =& \frac{2(m_{\rm Kerr})_{0}}{m_{2}} \frac{b}{(R)_{0}}
- \frac{1}{b^{2}} (\bar{\cal E}_{kl})_{0} (X^{k})_{0} (X^{l})_{0}
\,, \cr
(g^{\rm IZ}_{0i})_{2} =& \frac{1}{3b^{2}} \frac{(X_{i})_{0}}{(R)_{0}} (\bar{\cal E}_{kl})_{0} (X^{k})_{0} (X^{l})_{0}
\cr &
+ \frac{2}{3b^{2}} (R)_{0} (\bar{\cal E}_{ik})_{0} (X^{k})_{0}
\,, \cr
(g^{\rm IZ}_{ij})_{2} =& \biggl( \frac{2(m_{\rm Kerr})_{0}}{m_{2}} \frac{b}{(R)_{0}}
- \frac{1}{3b^{2}} (\bar{\cal E}_{kl})_{0} (X^{k})_{0} (X^{l})_{0} \biggr) \delta_{ij}
\cr &
- \frac{2}{3b^{2}} (\bar{\cal E}_{ij})_{0} (R)_{0}^{2}\,.
\label{eq:IZ_bz}
\end{align}
Here, the electric ${\cal E}_{kl}$ tidal tensor components
are related to the parameters $z_{R,m}$ and $z_{I,m}$ as
\begin{align}
{\cal E}_{XX} &= -\frac{1}{8} z_{R,-2}-\frac{1}{4} z_{R,0}-\frac{1}{8} z_{R,2} \,, 
\cr
{\cal E}_{XY} &= -\frac{1}{8} z_{I,-2}+\frac{1}{8} z_{I,2} \,,
\cr
{\cal E}_{XZ} &= -\frac{1}{4} z_{R,-1}-\frac{1}{4} z_{R,1} \,,
\cr
{\cal E}_{YY} &= \frac{1}{8} z_{R,-2}-\frac{1}{4} z_{R,0}+\frac{1}{8} z_{R,2} \,, 
\cr
{\cal E}_{YZ} &= -\frac{1}{4} z_{I,-1}+\frac{1}{4} z_{I,1} \,,
\cr
{\cal E}_{ZZ} &= \frac{1}{2} z_{R,0} \,,
\end{align}
where ${\cal E}_{XX}+{\cal E}_{YY}+{\cal E}_{ZZ}=0$,
and ${\cal E}_{kl}$ is expanded as
\begin{align}
{\cal E}_{kl} &= \frac{m_{2}}{b^{3}} (\bar{\cal E}_{kl})_{0} + {\cal{O}}(v^{3})
\,.
\end{align}
Since the magnetic tidal tensor components, ${\cal B}_{kl}$,
is higher order than ${\cal E}_{kl}$, we ignore them
when we discuss the matching up to $O[(m_{2}/b)^1]$, and  
$(\bar{\cal E}_{ij})_{0}$ is written as
\begin{align}
(\bar{\cal E}_{ij})_{0} &= \delta_{ij} - 3 \hat \beta_i \hat \beta_j \,.
\label{eq:bEij}
\end{align}
We are using the notation
$\hat \beta^\alpha = \{0,\,\cos\omega t,\,\sin\omega t,\,0 \}$
above.

%%%%%%%%%%%%%%%%%%%%
\subsection{Matching Calculation}
%%%%%%%%%%%%%%%%%%%%

We presented the formal expression of the asymptotic matching in Sec.~\ref{sec:AM},
then the IZ and NZ metrics in the BZ in Sec.~\ref{sec:expIZNZ}.
Using the results from Sec.~\ref{sec:expIZNZ}, 
we calculate the coordinate transformation for the asymptotic matching.
This consists of solving
Eq.~\eqref{eq:coord_transf} order by order
to $O[(m_{2}/b)^{1}]$
with respect to $(m_{2}/b)^{1/2}$.

%%%%%%%%%%
\subsubsection{Zeroth-order matching: $O[(m_{2}/b)^0]$}

At zeroth order, we have the matching equation
\begin{align}
(g^{\rm NZ}_{\alpha \beta})_{0} &= (A_{\alpha}{}^{\gamma})_{0} (A_{\beta}{}^{\delta})_{0} 
(g^{\rm IZ}_{\gamma \delta})_{0}
\,,
\end{align}
with $A_{\alpha}{}^{\beta} = \partial_{\alpha} X^{\beta}$.
Using $(g^{\rm NZ}_{\alpha \beta})_{0} = (g^{\rm IZ}_{\alpha \beta})_{0} = \eta_{\alpha \beta}$,
and taking into account the position of BH1, the zeroth order
coordinate transformation is given by
\begin{align}
(X^{\alpha})_{0} = x^{\alpha} - \frac{m_{2}}{M} b \,\hat \beta^{\alpha} = \tilde x^{\alpha} \,.
\label{eq:0thCT}
\end{align}
We also understand $(r_1^i)_0 = \tilde x^{i}$.
Here, it is noted that $\hat \beta^{\alpha}$ has a time dependence, i.e.,
\begin{align}
\partial_t (X^{\alpha})_{0} = \hat t^{\alpha} - \frac{m_{2}}{M} b \,\omega \,\hat \nu^{\alpha} 
= \hat t^{\alpha} - \sqrt{\frac{m_2}{b}} \sqrt{\frac{m_{2}}{M}} \,\hat \nu^{\alpha} \,,
\label{eq:corrT}
\end{align}
where $\hat t^\alpha=\{1,\,0,\,0,\,0\}$ and $\hat \nu^\alpha = \{0,\,-\sin\omega t,\,\cos\omega t,\,0 \}$.
The last term in the above equation creates the difference 
between Ref.~\cite{Gallouin:2012kb} and this paper.

%%%%%%%%%%
\subsubsection{First-order matching: $O[(m_{2}/b)^{1/2}]$}

At first order, the matching equation becomes
\begin{align}
(g^{\rm NZ}_{\alpha \beta})_{1} =&
(A_{\alpha}{}^{\gamma})_{0} (A_{\beta}{}^{\delta})_{0} (g^{\rm IZ}_{\gamma \delta})_{1}
\cr &
+ 2\,(A_{(\alpha}{}^{\gamma})_{1} (A_{\beta)}{}^{\delta})_{0} (g^{\rm IZ}_{\gamma \delta})_{0} \,,
\end{align}
where $T_{(\alpha\beta)}$ denotes symmetrization about two indices.
Using $(g^{\rm NZ}_{\alpha \beta})_{1} = (g^{\rm IZ}_{\alpha \beta})_{1} = 0$,
$(g^{\rm IZ}_{\gamma \delta})_{0} = \eta_{\alpha \beta}$,
$\partial_i (X^{\alpha})_{0} = \delta_{i}{}^{\alpha}$
and Eq.~\eqref{eq:corrT}, the above equation is written as
\begin{align}
(A_{(\alpha \beta)})_{1} 
+ \sqrt{\frac{m_{2}}{M}} \,\hat t_{(\alpha} \hat \nu_{\beta)} = 0 \,.
\end{align}
One of the solutions can be obtained as
\begin{align}
(X^{\alpha})_{1} &= - \sqrt{\frac{m_{2}}{M}} \tilde y_{\rm C} \, \hat t^{\alpha} \,,
\end{align}
where $\tilde y_{\rm C} = \hat \nu_{i} \tilde x^{i} = \hat \nu_{\alpha} \tilde x^{\alpha}$.
We also use the notation
$\tilde x_{\rm C} = \hat \beta_{i} \tilde x^{i} = \hat \beta_{\alpha} \tilde x^{\alpha}$
in the following analysis. $\tilde x_{\rm C}$ and $\tilde y_{\rm C}$
are the coordinates centered on BH1 that
are co-rotating with the binary.

%%%%%%%%%%
\subsubsection{Second-order matching: $O[(m_{2}/b)^1]$}

In the above leading and first order analysis, we have derived
\begin{align}
(X^{\alpha})_{\{1\}} &= \tilde x^{\alpha} 
- \sqrt{\frac{m_2}{b}} \sqrt{\frac{m_{2}}{M}} \, \tilde y_{\rm C} \, \hat t^{\alpha} \,,
\end{align}
where $\{1\}$ denotes the leading $+$ first order quantity.
At second order, we have a formal expression for the matching as
\begin{align}
(g^{\rm NZ}_{\alpha \beta})_{\{2\}} =& (A_{\alpha}{}^{\gamma})_{\{2\}} 
(A_{\beta}{}^{\delta})_{\{2\}} (g^{\rm IZ}_{\gamma \delta})_{\{2\}} \,.
\end{align}
Again, $\{2\}$ means the leading $+$ first order $+$ second order quantity,
and $(A_{\alpha}{}^{\gamma})_{\{2\}}$ is written by
\begin{align}
(A_{\alpha}{}^{\gamma})_{\{2\}} =& \delta_{\alpha}{}^{\gamma} 
+ \sqrt{\frac{m_2}{b}} \biggl[ \sqrt{\frac{m_{2}}{M}} \,\hat t_{\alpha} \hat \nu^{\gamma}
- \sqrt{\frac{m_{2}}{M}} \, \hat \nu_{\alpha} \hat t^{\gamma} \biggr]
\cr &
+ \frac{m_2}{b} 
\biggl[ - \frac{1}{b} \left(\tilde x_{\rm C} + \frac{m_2}{M} b \right) \, \hat t_{\alpha} \hat t^{\gamma}
+ (\partial_{\alpha} X^{\gamma})_{2} \biggr] \,.
\end{align}
Finding the solution for $(X^{\gamma})_{2}$ is the remaining task to complete.

Using the explicit expression of
\begin{widetext}
\begin{align}
[(A_{\alpha}{}^{\gamma})_{(2)} (A_{\beta}{}^{\delta})_{(2)} \eta_{\gamma \delta}]_{(2)} =
\eta_{\alpha \beta} 
+ \frac{m_2}{b} 
\biggl[
- \frac{2}{b} \tilde x_{\rm C} \hat t_{\alpha} \hat t_{\beta} 
- \frac{m_2}{M} \hat t_{\alpha} \hat t_{\beta}
- \frac{m_{2}}{M} \hat \nu_{\alpha} \hat \nu_{\beta}
+ 2 (A_{(\alpha \beta)})_{2}
\biggr] \,,
\end{align}
and Eq.~\eqref{eq:bEij} for $(\bar{\cal E}_{kj})_{0}$, we may solve
\begin{align}
2(A_{(\alpha \beta)})_{2} =&
\biggl[ \biggl(2-\frac{2}{b} \tilde x_{\rm C} \biggr) \Delta_{\alpha \beta}
+ \frac{2}{b} \tilde x_{\rm C} \hat t_{\alpha} \hat t_{\beta} 
+ \frac{m_2}{M} \hat t_{\alpha} \hat t_{\beta}
+ \frac{m_{2}}{M} \hat \nu_{\alpha} \hat \nu_{\beta} \biggr]
+ \biggl[ \delta_{\alpha}^{i} \delta_{\beta}^{j} \frac{2}{b^{2}}
\Bigl( \tilde x_{\rm C}^2 \delta_{ij} - (r_1)_0^2 \hat \beta_{i} \hat \beta_{j} \Bigr) \biggr]
\cr &
+ \biggl[ ( \delta_{\alpha}^{i} \hat t_\beta + \delta_{\beta}^{i} \hat t_\alpha ) \frac{1}{3b^{2}}
\Bigl( 3 (r_1)_0 \tilde{x}_{i} 
- \frac{3}{(r_1)_{0}} \tilde x_{\rm C}^2 \tilde x_{i} 
- 6 (r_1)_{0} \tilde x_{\rm C} \hat \beta_{i} \Bigr) \biggr]
\,.
\label{eq:for2ndCT}
\end{align}
\end{widetext}

The second order coordinate transformation $(X_{\alpha})_{2}$ is derived
as follows. From the first bracket in Eq.~\eqref{eq:for2ndCT},
we obtain a particular solution,
\begin{align}
(X_{\alpha})_{2,p1} =& 
\left( 1+\frac{m_2}{2 M} \right) (\tilde{x}^{\beta} \hat t_{\beta}) \hat t_{\alpha} 
+ \left( 1-\frac{\tilde x_{\rm C}}{b} \right) \Delta_{\alpha i} \tilde{x}^{i}
\cr &
+ \frac{\Delta_{ij} \tilde{x}^{i} \tilde{x}^{j}}{2b} \hat \beta_{\alpha}
+ \frac{m_2}{2 M} \tilde y_{\rm C} \hat \nu_{\alpha} \,,
\end{align}
and from the second bracket, a particular solution is
\begin{align}
(X_{\alpha})_{2,p2} &= - \frac{1}{b^{2}} \left( (r_1)_0^2 \tilde x_{\rm C} \hat \beta_{i}
 - \tilde x_{\rm C}^{2} \tilde{x}_{i} \right) \delta_{\alpha}^{i} \,.
\end{align}
The third bracket gives a particular solution,
\begin{align}
(X_{\alpha})_{2,p3} &= \frac{1}{3b^{2}} \left( (r_1)_0^3
- 3 \tilde x_{\rm C}^{2} (r_1)_0 \right) \hat t_\alpha \,.
\end{align}
Finally, combining the above three particular solutions,
the coordinate transformation is written as
\begin{align}
(X_{\alpha})_{2} =&
(X_{\alpha})_{2,p1} + (X_{\alpha})_{2,p2} + (X_{\alpha})_{2,p3} \,,
\end{align}
where we have ignored the homogeneous solution which is required
in higher order matching.
This means that the resultant coordinate transformation is not unique.
The series expansion of $(X^{\alpha})_{0}+\sqrt{m_2/b}(X^{\alpha})_{1}+(m_2/b)(X^{\alpha})_{2}$
with respect to $t/b \ll 1$ gives the same coordinate transformation
as obtained in Ref.~\cite{Gallouin:2012kb}.

In practice, we use the following explicit expressions for the coordinate
transformation.  Using the PN orbital phase evolution $\omega t = \phi = \phi(t)$
and the PN evolution of the orbital separation $b = r_{12} = r_{12}(t)$,
and introducing the notations $\tilde x^{\alpha}=\{t,\,\tilde
x,\,\tilde y,\,z\}$ and $\tilde r_1=\sqrt{\tilde x^{i}\tilde x_{i}}(=(r_1)_0)$, 
\begin{widetext}
\begin{align}
\label{eq:CTforLTE}
T &= t-\sqrt {{\frac {m_2}{r_{12}}}}\sqrt {{\frac {m_2}{M}}}{\tilde y_{\rm C}}
+\frac{m_2}{{r_{12}}}\, \left( \frac{1}{3}\,{\frac {{{\tilde r_1}}^{3}-3\,{{\tilde x_{\rm C}}}^{2}{\tilde r_1}}{{r_{12}}^{2}}}
- \left( 1+\frac{1}{2}\,{\frac {m_2}{M}} \right) t \right) 
\cr
&= t-\sqrt {{\frac {m_2}{r_{12}}}}\sqrt {{\frac {m_2}{M}}}{\tilde y_{\rm C}}
+\frac{m_2}{{r_{12}}}\, \left( \frac{1}{3}\,{\frac {{{\tilde r_1}}^{3}-3\,{{\tilde x_{\rm C}}}^{2}{\tilde r_1}}{{r_{12}}^{2}}}
 \right) 
+ \frac{5}{384} \frac{(2M + m_2)(r_{12}^3-r_{12}(0)^3)}{M^2 m_1} \,,
\cr
X &= {\tilde x}+ \frac{m_2}{{r_{12}}}\, \left( -{\frac {{{\tilde r_1}}^{2}{\tilde x_{\rm C}}\,\cos \phi 
-{{\tilde x_{\rm C}}}^{2}{\tilde x}}{{r_{12}}^{2}}}+{\tilde x}\, \left( 1-{\frac {{\tilde x_{\rm C}}}{r_{12}}} \right) 
+\frac{1}{2}\,{\frac {{{\tilde r_1}}^{2}\cos \phi }{r_{12}}}
-\frac{1}{2}\,{\frac {m_2\,{\tilde y_{\rm C}}\,\sin \phi }{M}} \right) \,,
\cr
Y &= {\tilde y}+\frac{m_2}{{r_{12}}}\, \left( -{\frac {{{\tilde r_1}}^{2}{\tilde x_{\rm C}}\,\sin \phi 
-{{\tilde x_{\rm C}}}^{2}{\tilde y}}{{r_{12}}^{2}}}+{\tilde y}\, \left( 1-{\frac {{\tilde x_{\rm C}}}{r_{12}}} \right) 
+\frac{1}{2}\,{\frac {{{\tilde r_1}}^{2}\sin \phi }{r_{12}}}
+\frac{1}{2}\,{\frac {m_2\,{\tilde y_{\rm C}}\,\cos \phi }{M}} \right) \,,
\cr
Z &= z+\frac{m_2}{{r_{12}}}\, \left( {\frac {{{\tilde x_{\rm C}}}^{2}z}{{r_{12}}^{2}}}
+z \left( 1-{\frac {{\tilde x_{\rm C}}}{r_{12}}} \right)  \right) \,.
\end{align}
\end{widetext}
Here, some terms with $t$ in the $T$-component have been rewritten
via the rate of change of the orbital separation as in 
Ref.~\cite{Mundim:2013vca},
\begin{align}
\frac{t}{r_{12}} &= \int_0^t \frac{dt}{r_{12}} 
= \int_{r_{12}(0)}^{r_{12}} dr_{12} \left( \frac{dr_{12}}{dt} \right)^{-1}\frac{1}{r_{12}}
\cr 
&= - \int_{r_{12}(0)}^{r_{12}} dr_{12} \frac{5 r_{12}^2}{64 M^3 \eta} 
\cr &
= - \frac{5 (r_{12}^3-r_{12}(0)^3)}{192 M^3 \eta} \,,
\end{align}
where $r_{12}(0)$ is the initial orbital separation which we set in the
numerical calculation.

%%%%%%%%%%%%%%%%%%%%
\subsection{Global Metric}
%%%%%%%%%%%%%%%%%%%%

With the asymptotic matching of IZA (A$=1,\,2$) to the NZ in hand, 
we can stitch the IZ metric to the NZ metric 
(and similarly with the NZ to FZ) together 
via the proper transition functions in the BZ in Fig.~\ref{fig:zones} and Table~\ref{tab:zones}. 
These transition functions are specially selected 
to obey the Frankenstein theorems of Ref.~\cite{Yunes:2006mx}, 
and therefore will not introduce any error 
into the metric calculation that is larger 
than the error already generated in the individual zones. 
The global metric is then a weighted average
\begin{align}
g_{\mu\nu} =&
(1 - f_{\rm far})
\Bigl\{f_{\rm near} \bigl[f_{{\rm inner},1} \,g_{\mu\nu}^{\rm NZ} 
+(1 - f_{{\rm inner},1} ) \,g_{\mu\nu}^{\rm IZ1}\bigr]
\nonumber \\ &
+ (1 - f_{\rm near} )\bigl[f_{{\rm inner},2} \,g_{\mu\nu}^{\rm NZ} 
+(1 - f_{{\rm inner},2} ) \,g_{\mu\nu}^{\rm IZ2}\bigr]\Bigr\}
\nonumber \\ &
+ f_{\rm far} \,g_{\mu\nu}^{\rm FZ} \,,
\label{eq:wholemetric}
\end{align}
where the transition functions $f_{\rm far}$, $f_{\rm near}$, 
$f_{{\rm inner},1}$, and $f_{{\rm inner},2}$ are summarized 
in Appendix~\ref{app:trans_func}.

Here, it is noted that we have used various different type/order approximations
in the IZ, NZ, and FZ metrics, and the EOM.
Therefore, to choose the BZs, 
we need to take into account for the largest possible error which arises from the finite order
truncation in the approximations, for example, $O[(m_2/b)^{3/2}]$ in the IZ1-NZ BZ.
Using these BZs, we can obey the Frankenstein theorems of Ref.~\cite{Yunes:2006mx},
and avoid any unphysical behavior due to different approximations.

To demonstrate that the matching and the construction of the global metric do
not introduce any pathological behavior in the coordinate choice outside the
horizon, we show here the volume element, $\sqrt{-g}$, for the global metric,
which encodes, for example, the IZ metric in the PN harmonic coordinates, after
the coordinate transformation and transition function have been carried out.

\begin{figure}[!htb]
\begin{center}
\includegraphics[width=\columnwidth,clip=true]{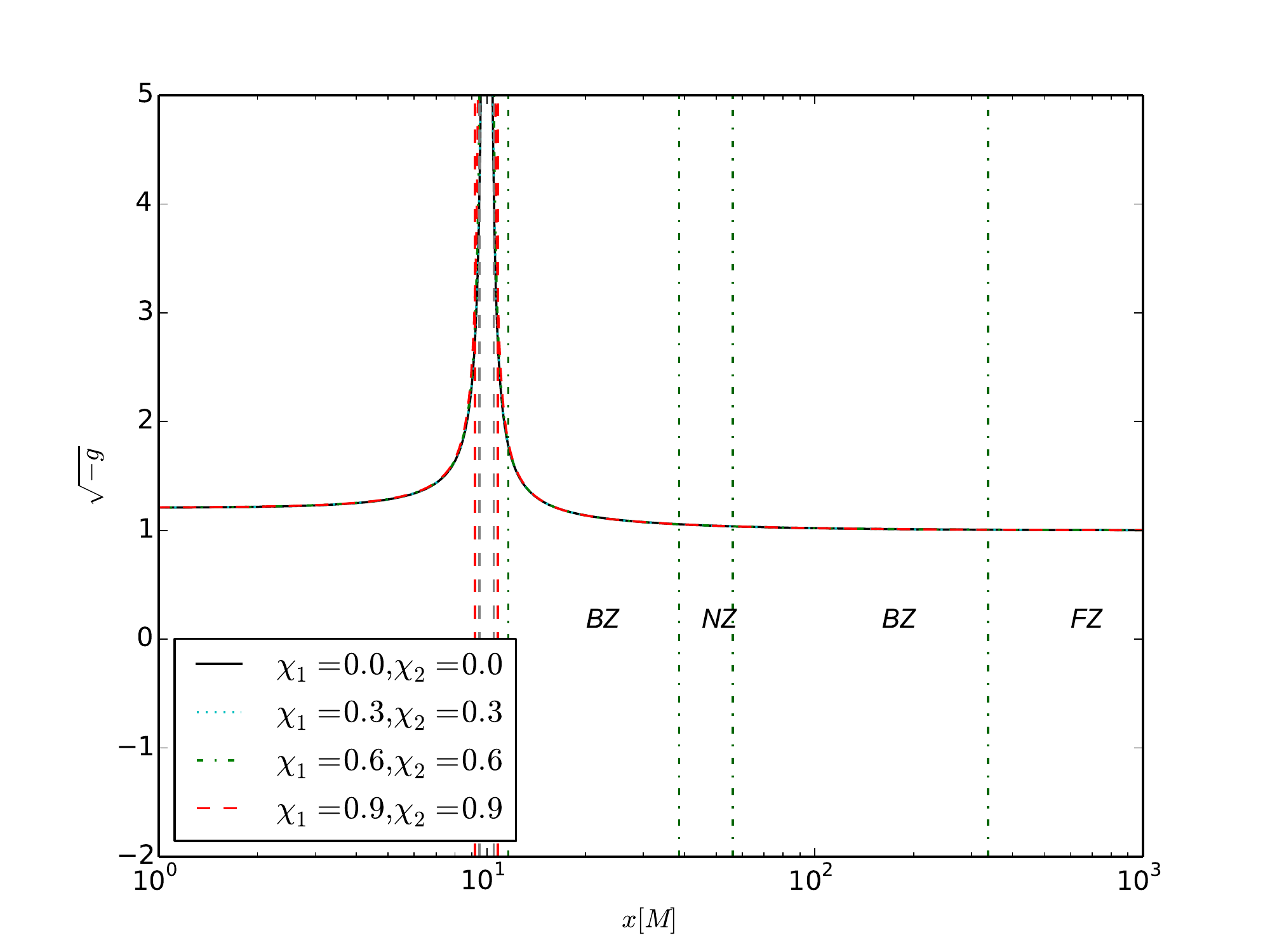}
\end{center}
\caption{
Comparison of the volume element, $\sqrt{-g}$, for the global metric for
differing values of the spin parameter $\chi$, where both black holes have spin
aligned with the orbital angular momentum of the binary.
}
\label{fig:detg}
\end{figure}

%%%%%%%%%%%%%%%%%%%%%%%%%%%%%%%%%%%%%%%%
\section{NUMERICAL ANALYSIS}\label{sec:numerical}
%%%%%%%%%%%%%%%%%%%%%%%%%%%%%%%%%%%%%%%%

To verify the correctness of the analytic metric approximating a spinning BHB spacetime,
we developed a battery of different and independent
tests to judge the quality of the analytic approximation
as in Ref~\cite{Mundim:2013vca}. We are mainly 
interested in identifying how much this analytic approximation deviates 
from the true solution to the Einstein's equations. In order to achieve
a reasonable, independent analysis of this approximation we resort then to the
computation of spacetime scalar invariants and their comparison against their 
expected values for vacuum spacetimes. While these analysis might not be 
sufficient to judge all aspects of this new metric it is definitely necessary
to assess its overall quality especially when compared against other analytic, 
approximative metrics.

We summarize in the following subsections which scalar invariants
we have used in our analysis. We then present the results for this analytic metric. 
Discussion on how these scalar invariants 
are computed in our codes is left to the end of this section~\ref{sec:Numerical_methods}.

%%%%%%%%%%%%%%%%%%%%
\subsection{Ricci Scalar}
%%%%%%%%%%%%%%%%%%%%

Using this global approximate metric, we calculate the Einstein tensor $G_{\mu
\nu}$, the Ricci tensor $R_{\mu \nu}$, the Ricci Scalar $R$ and the
relative Kretschmann invariant $K$ (see below) to test the validity of the
approximate metric.  If the BHB spacetime constructed is a valid vacuum
solution, then it naturally must satisfy the ten  Einstein equations in vacuum,
$R_{\mu \nu} = 0$.  Therefore, any deviations from $ R = g_{\mu \nu} R^{\mu
\nu} = 0$ can be interpreted as a measure of the violation to the Einstein
equations that the global metric has incurred by the approximate construction.  In the
following analysis, we use the sign conventions laid out in Ref.~\cite{MTW}
(see also Wald's ``General Relativity''~\cite{Wald:1984rg})
for all different geometric quantities entering the computation of Ricci scalar.
Note also that we have used projections of the
Ricci tensor along the hyperspace normal and into the time slices to compute
the Hamiltonian and Momentum constraints, which are consistent with the Ricci 
scalar analysis we present.

%%%%%%%%%%%%%%%%%%%%
\subsection{Relative Kretschmann}
%%%%%%%%%%%%%%%%%%%%

In principle, it is possible to construct many invariants
for the BHB problem. Here we present a new concept that we can use to
evaluate the validity of the approximate, analytic metric.  One of the pitfalls
of using a quantity such as the Ricci scalar in analysis of the violation to
the Einstein equations is that it is not a normalized quantity. The Ricci
scalar can be large without bound, and it is therefore difficult to assign
meaning to a numerical quantity in the Ricci scalar without a scale to compare 
our results with.  
The only scale that the Ricci scalar provides in vacuum
is how far it deviates from zero.  A Ricci scalar value of
$10^{-9}$ is better than a value of $10^{-6}$, but that is all that we can
really say about it. If we wish to use this as an assessment of the error,
it becomes difficult to assign meaning to the results if they are far from 
zero. 

It would therefore be desirable to have a quantity that both measured
the violation of the Einstein equations and provides a scale, so values could 
be compared directly and we can easily interpret them.

For this purpose, we introduce an invariant that can still give us a measure
of the violation to the Einstein equations and has the
added benefit of being normalized: the relative Kretschmann curvature scalar.

We start with the definition of the Weyl tensor $C_{\mu \nu \rho \delta}$ from
Ref.~\cite{Wald:1978vm},
\begin{align}
R_{\mu \nu \rho \delta} =& C_{\mu \nu \rho \delta} 
+ (g_{\mu [\rho}R_{\delta ] \nu} - g_{\nu [ \rho} R_{\delta ] \mu}) 
- \frac{1}{3} R g_{\mu [ \rho} g_{\delta ] \nu} \,.
\end{align}
From here, we contract the Weyl tensor with itself, eventually yielding
Kretschmann curvature scalar: 
\begin{align}
R_{\mu \nu \rho \delta} R^{\mu \nu \rho \delta} = 
C_{\mu \nu \rho \delta} C^{\mu \nu \rho \delta}  + 2 R_{\mu \nu} R^{\mu \nu} - \frac {1}{3} R^2 \,.
\label{eq:RRCC}
\end{align}
We now can say that if the solution is exact, we know that this contraction of
the Riemann tensor should be equal exactly to the contraction of the Weyl
tensor. In exact solutions to Einstein's equations, contraction of the Riemann
tensor and the Ricci scalar are both zero.  Therefore, we can define a relative
Kretschmann from the remainder: 
\begin{align}
K_{\rm rel} = \bigg|  \frac{ 2 R_{\mu \nu} R^{\mu \nu} - R^{2}/3}
{R_{\mu \nu \rho \delta} R^{\mu \nu \rho \delta}} \bigg| \,,
\label{eq:relK}
\end{align} 
which is the remainder from the exact vacuum solution normalized by the
Kretschmann invariant $R_{\mu \nu \rho \delta} R^{\mu \nu \rho \delta}$.
We expect that this value will be less than one anywhere in the global spacetime
for small violations from the vacuum spacetime.
For larger violations, it may be possible to have $K_{\rm rel}>1$.
This is because there is no constraint for the energy-momentum tensor
which is converted from the Ricci violation.
We can now use this as a measure of the exactness of the solution,
and plot the residual that
we obtain to get an idea what the relative violation is to the true (exact)
solution. Essentially this normalization introduces a scale to which we can
compare the errors our approximation produces, giving us the desirable feature of 
having a direct way achieve this task in our spacetime.

%%%%%%%%%%%%%%%%%%%%
\subsection{Accuracy of the Global Metric: the Ricci Scalar}\label{sec:accuracy}
%%%%%%%%%%%%%%%%%%%%

\begin{figure}[!htb]
\begin{center}
\includegraphics[width=\columnwidth,clip=true]{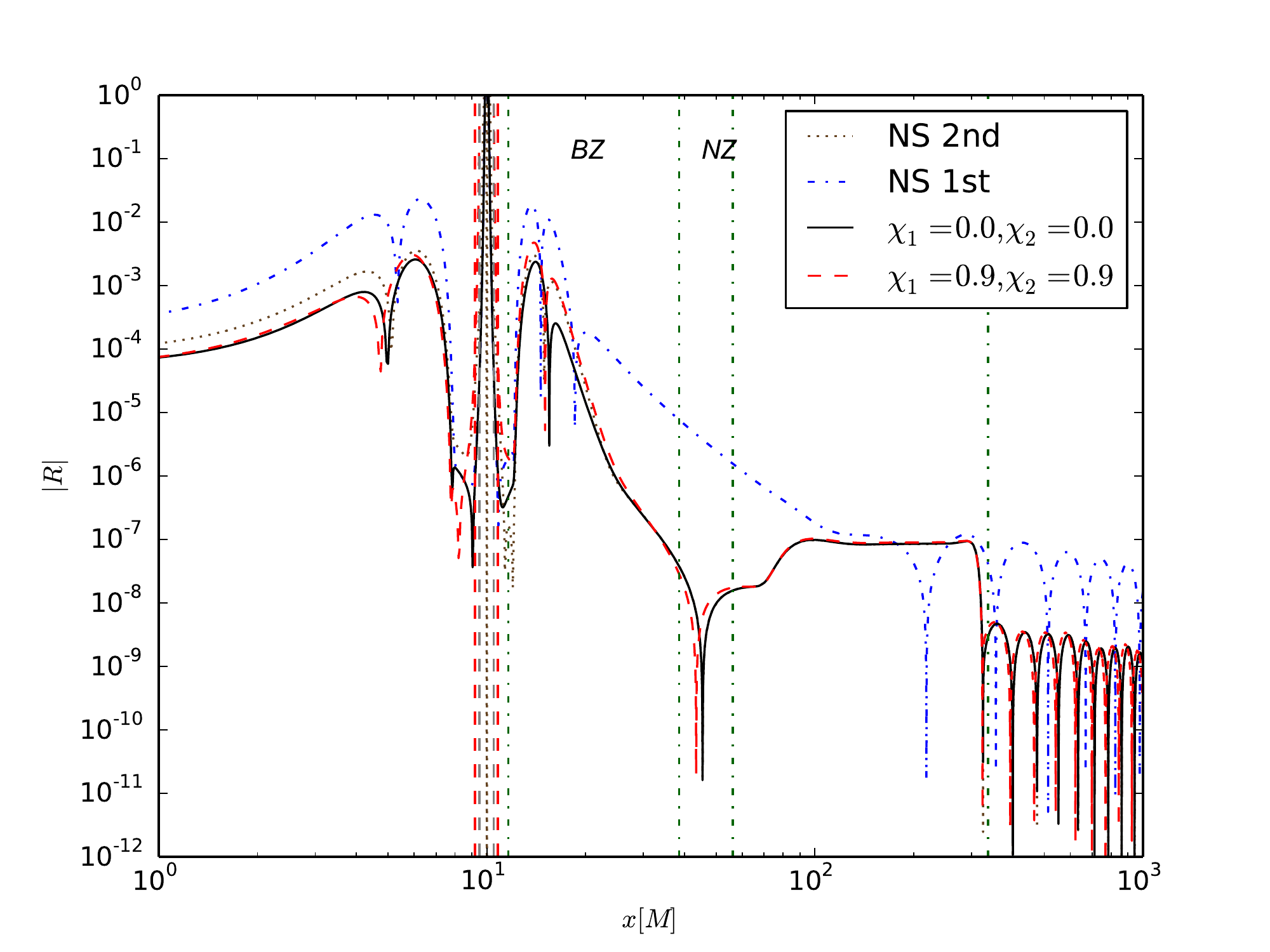}
\end{center}
\caption{
The absolute value of the Ricci scalar along the $x$ coordinate, at an initial separation of 
$20M$.  We compare here
the violations of the Ricci scalar for the spinning BHB metric against the
non-spinning (NS) BHB first and second order metrics~\cite{Mundim:2013vca}. It is
interesting to note that the new spinning BHB metric has violations of the same
order of magnitude as the previous non-spinning BHB metric.  In this plot and
the following plots, the horizon is denoted by the grey dashed vertical line,
discussed in more detail in Fig.~\ref{fig:aligned_zoom}. The ISCO is the orange
dashed vertical line (see Appendix~\ref{app:ISCO} for details).  The green
dash-dotted vertical lines indicate the boundaries of the different zones, and
are consistent for all of the subsequent plots. It is good to note here that
the zone boundaries do change for differing spins, though not by much, so here
we picked the fiducial zone boundary for the $\chi_i = 0.9$ (highly spinning) and aligned
case.
}
\label{fig:Ricci_nospin_vs_spin}
\end{figure}

In Ref.~\cite{Mundim:2013vca}, we showed how the violations of the
Ricci scalar change as we increase the order of approximation for an equal mass
non-spinning BHB spacetime.  Fortunately as expected from the analytical point
of view, those violations became smaller everywhere as we went from a first
order metric (with the quadrupole (IZ)-1PN (NZ) matching)
to a second order metric (with the octupole (IZ)-2PN (NZ) matching).
We reproduce that result here in
Fig.~\ref{fig:Ricci_nospin_vs_spin}. Our work in this paper is the first step
towards higher order of approximation for spinning BHBs. As shown in Sec.~\ref{sec:AM},
the matching for the spacetime construction is
first order with the quadrupole order for the IZ and 1PN order for the NZ.
Since we do not have at the moment a higher order spinning BHB
spacetime to compare to, we use the first and second order metrics for
non-spinning BHB  as a reference for the spinning metric. The idea is make sure
that the spinning BHB metric does not introduce any larger violations of the
Ricci scalar than what we have already seen in the non-spinning case. As we can
see in Fig.~\ref{fig:Ricci_nospin_vs_spin} this is fortunately the case. The
first order matched spinning BHB metric results into Ricci violations that
follow most closely the second order violations of the non-spinning metric for
regions far away from the BHs and lies in between first and second order cases
for regions closer to the BHs. The reason we obtain much better results for the
first order spinning BHB metric than for the first order non-spinning BHB
metric is that we are indeed using higher order spinning metric components here
while keeping the matching to first order only. The rationale is that while we
would like to have a consistent order counting in this work and a future one
for second order one, we can already take advantage of higher order metric
pieces with smaller Ricci violations right now for our upcoming gas and MHD
simulations.

\begin{figure}[!htb]
\begin{center}
\includegraphics[width=\columnwidth,clip=true]{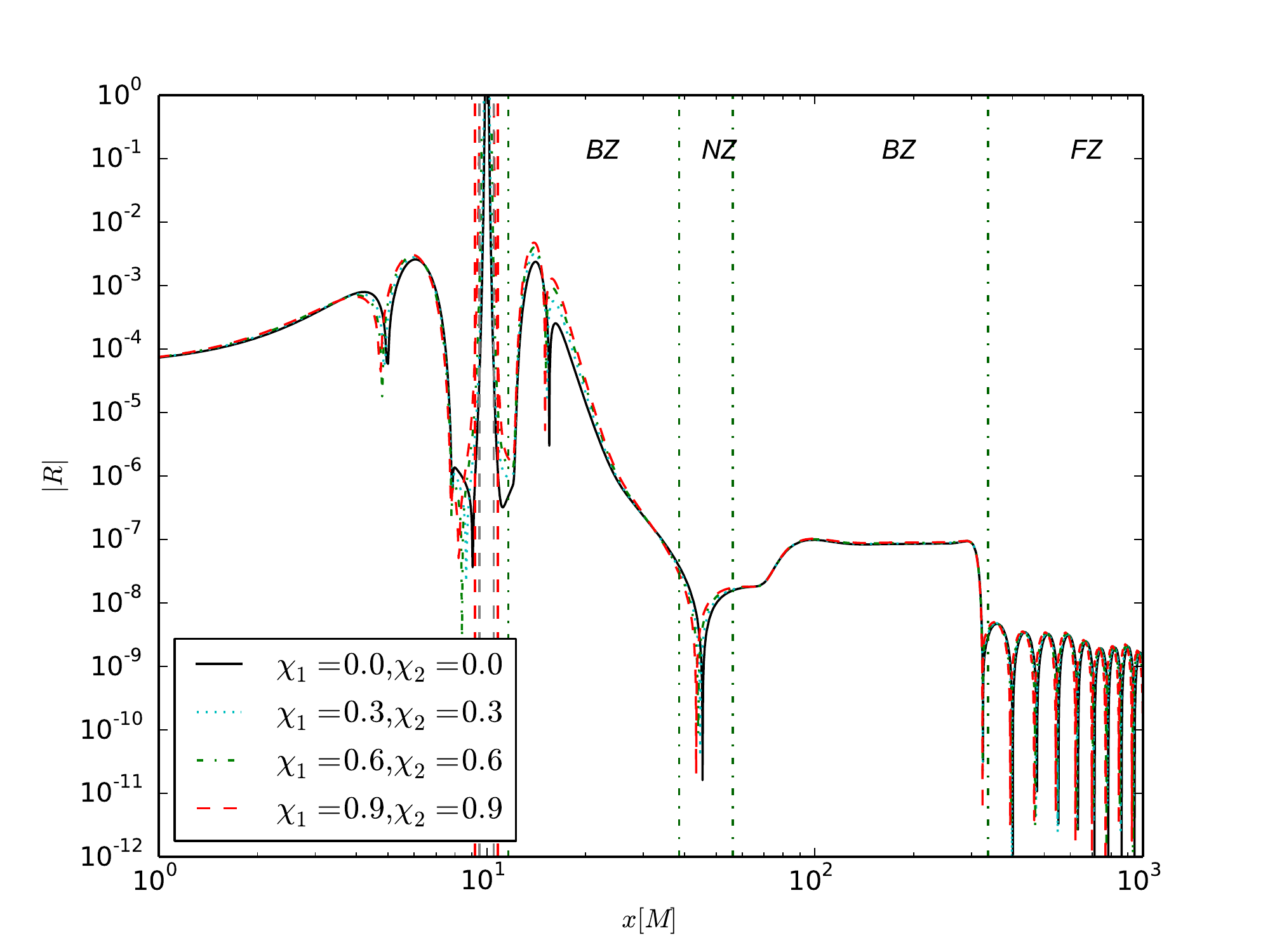}
\end{center}
\caption{The absolute value of the Ricci scalar along $x$, at an initial
separation of $20M$.  The spin parameter is varied in this plot. We see that
there is little qualitative variation in changing $\chi_i$. 
}
\label{fig:Ricci_aligned}
\end{figure}

\begin{figure}[!htb]
\begin{center}
\includegraphics[width=\columnwidth,clip=true]{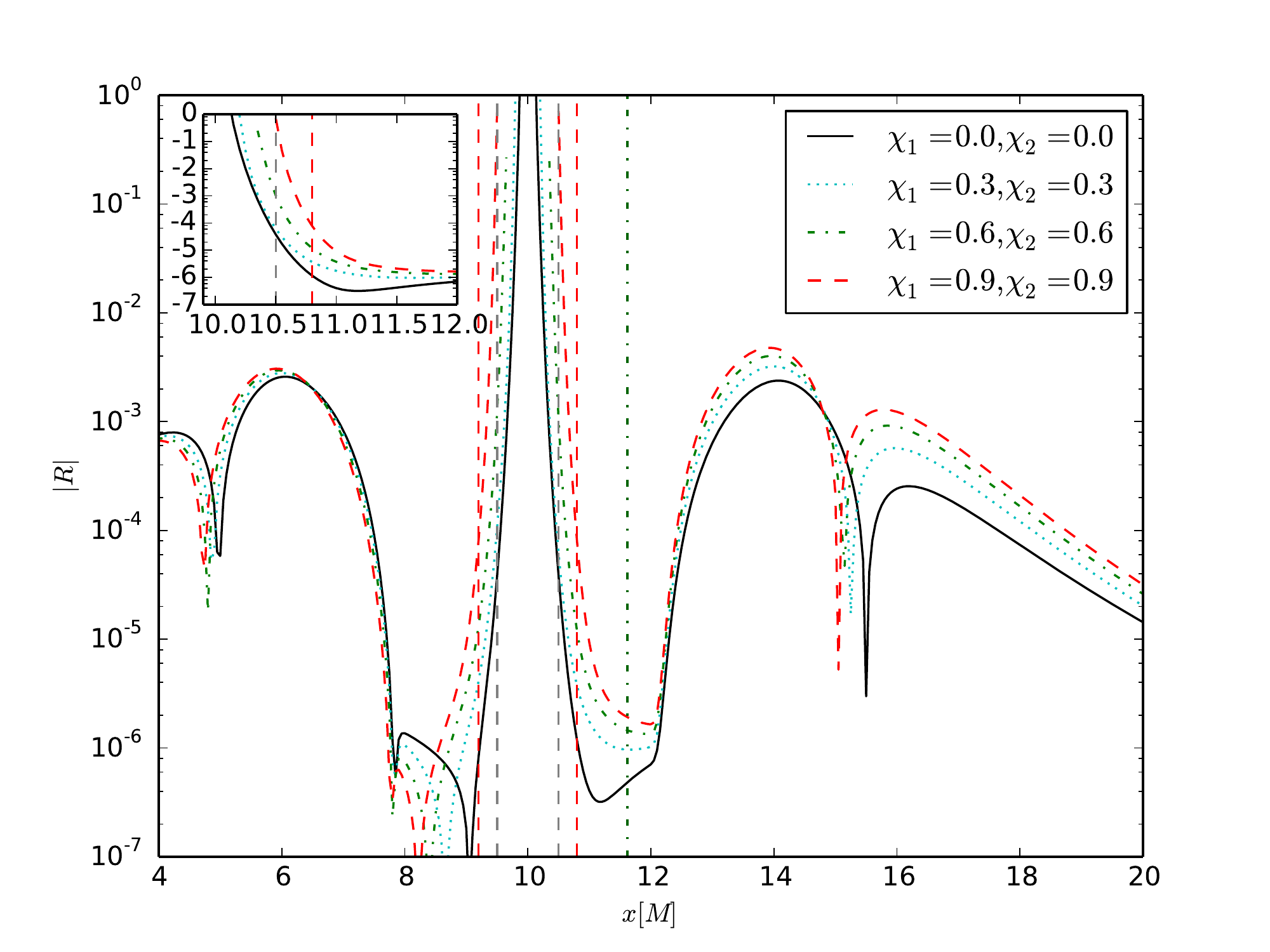}
\end{center}
\caption{The absolute value of the Ricci scalar along $x$, for aligned spins,
zoomed in on the inner region near the horizon. The horizon is denoted by the
grey dashed vertical line, roughly at a position of $x_{h} = x_{\rm BH,1} \pm M/2$
on the $x$-axis, 
and is easier to distinguish here than in the previous plots. The ISCO
is the orange dashed vertical line (see Appendix~\ref{app:ISCO} for details).
The inset shows the behavior close to the horizon.
}
\label{fig:aligned_zoom}
\end{figure}

As we increase the spin parameter value, $\chi_i$, from its non-spinning value,
$\chi_i=0$, to a very large spinning configuration, $\chi_i=0.9$, we observe
very little variations in the Ricci violations in the NZ and FZ.
That indicates the perturbation by spin addition to the system is small
(see Fig.~\ref{fig:Ricci_aligned}). As we zoom into the NZ, 
Fig.~\ref{fig:aligned_zoom}, the differences in violation amongst all spinning
cases become more evident. Only at a small spacetime volume between the 
horizon location and a radius set by an ISCO for an individual BH do these violation
differences span more than one order of magnitude. While of great importance
to accurately describe the spacetime in the vicinity of a BH, it is not 
so crucial in determining particle or gas dynamics since they are expected
to follow unstable circular orbits and accrete into the BH. We hope in 
a future work to improve on these violation differences between the spinning
cases by introducing a second order asymptotic matching between the IZ and NZ.

\begin{figure}[!htb]
\begin{center}
\includegraphics[width=\columnwidth,clip=true]{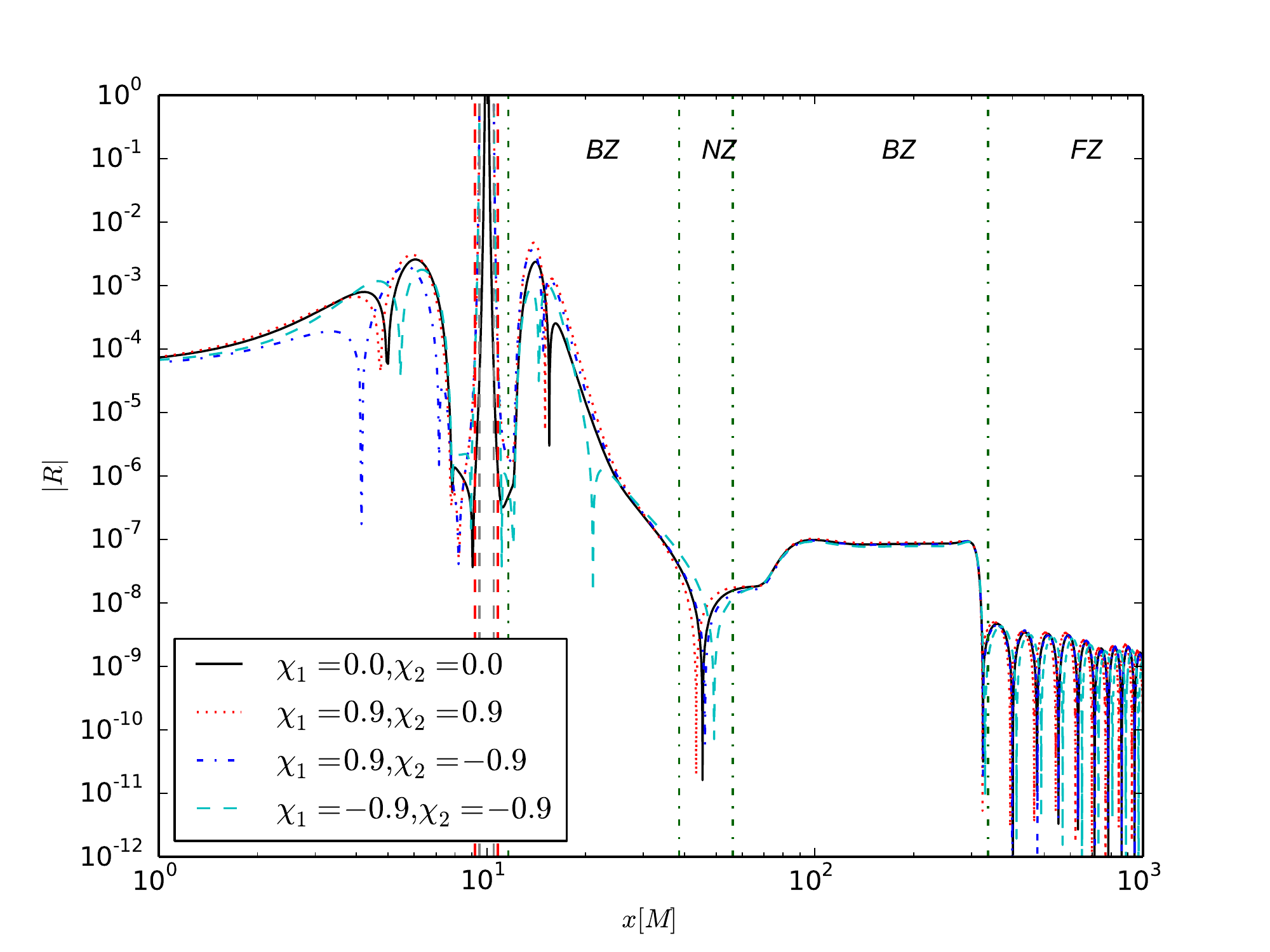}
\end{center}
\caption{The absolute value of the Ricci scalar along $x$, at an initial
separation of $20M$.  This plot shows the spin parameter $\chi_i$ varied for
anti-aligned spins.}
\label{fig:anti-aligned}
\end{figure}

\begin{figure}[!htb]
\begin{center}
\includegraphics[width=\columnwidth,clip=true]{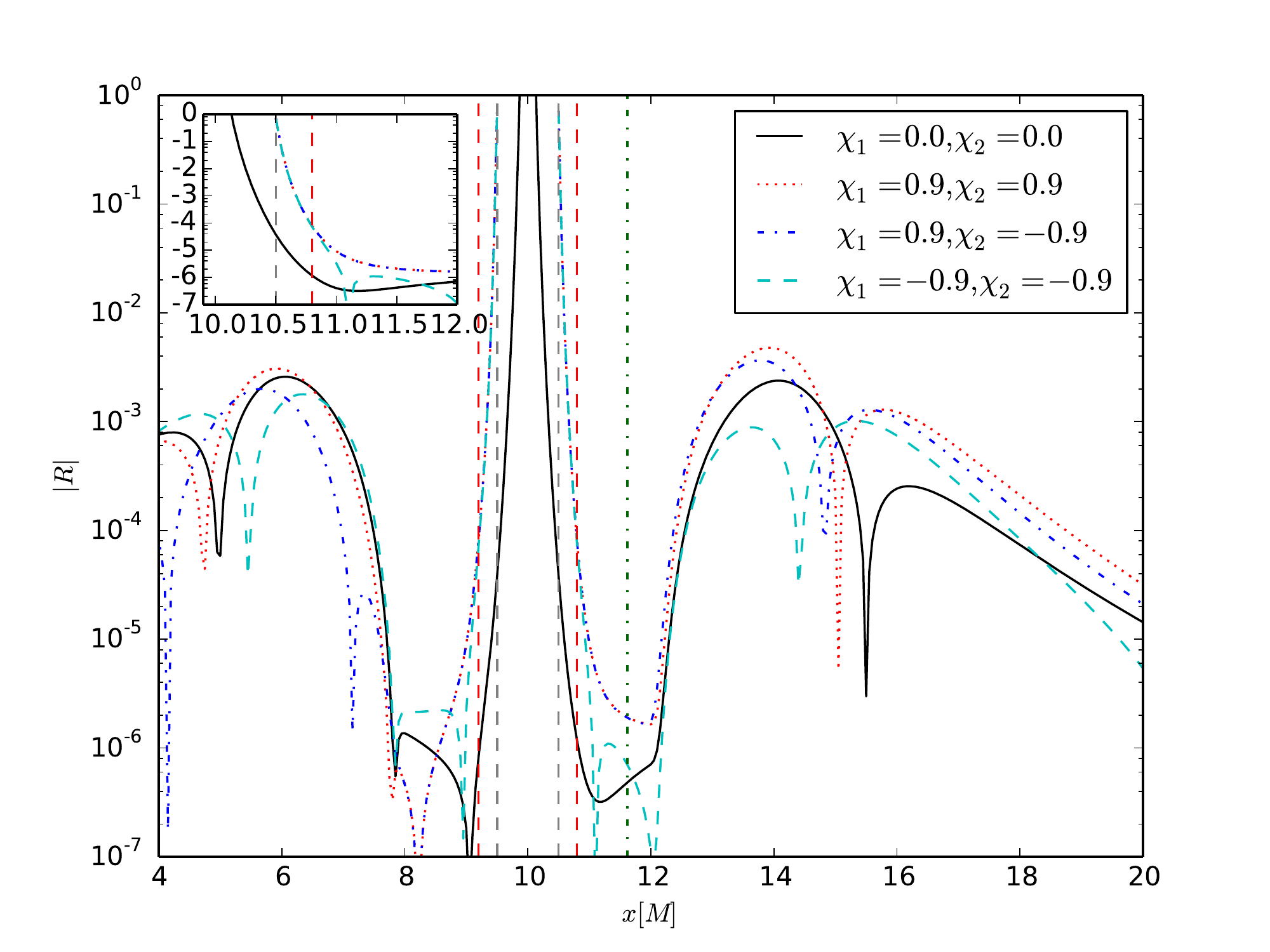}
\end{center}
\caption{The absolute value of the Ricci scalar along $x$, 
zoomed in near the BH to show the violation near
the horizon.  This plot shows the spin parameter $\chi_i$ varied for anti-aligned
spins. The inset shows the behavior close to the horizon.
}
\label{fig:anti-aligned_zoom}
\end{figure}

Next we exploit the effects of spin anti-alignment with the orbital angular
momentum. We fix our attention to the large spinning case, $|\chi_i| = 0.9$.
Again very little variation amongst the aligned, anti-aligned and the zero-sum
cases is observed in the NZ and FZ (see Fig.~\ref{fig:anti-aligned}).
As we zoom into the IZ, Fig.~\ref{fig:anti-aligned_zoom}, we can 
distinguish better among the cases, but none of them differ from each other
more than two orders of magnitude at ISCO locus.

\begin{figure}[!htb]
\begin{center}
\includegraphics[width=\columnwidth,clip=true]{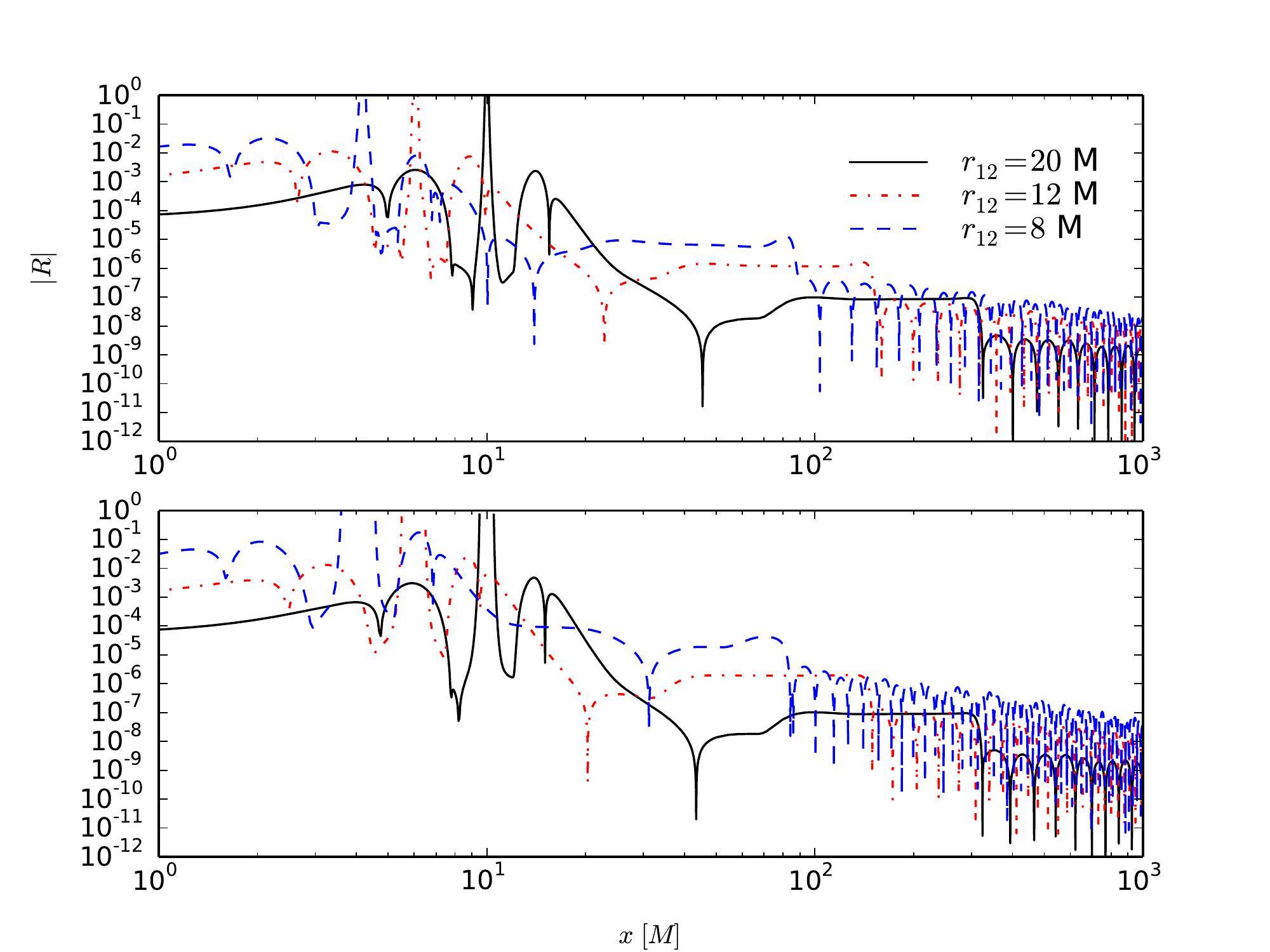}
\end{center}
\caption{The absolute value of the Ricci scalar along $x$, at an initial
separation of $20M$, $12M$, and $8M$, for spins $\chi_1 = \chi_2 = 0.0$ (top)
and $\chi_1 = \chi_2 = 0.9$ aligned  (bottom). Note that the violation increases
smoothly as we decrease in orbital separation. This gives good indication that the 
dynamics is not introducing any spurious error into the metric.
}
\label{fig:chi_evolution}
\end{figure}

Finally we show snapshots of the Ricci scalar as the binary evolves in time,
starting from a separation of $r_{12}=20M$ up until $r_{12}=8M$ roughly.
We were careful to pick the instants of time when the BHs cross the x axis
so that a comparison would be meaningful. As the separation decreases, the
perturbation parameters become larger and larger leading to a poorer 
approximation of the spacetime. As expected then, the violations of the 
Ricci scalar increases with evolution time or decreasing binary separation
as Fig.~\ref{fig:chi_evolution} shows for the aligned $\chi_i=0.9$ case.
The take home lesson from that figure is that the violations do increase
with time, but in a orderly and smooth fashion.

%%%%%%%%%%%%%%%%%%%%
\subsection{Accuracy of the Global Metric: the Relative Kretschmann}\label{sec:RKaccuracy}
%%%%%%%%%%%%%%%%%%%%

Here we plot the accuracy of the metric with respect to the relative Kretschmann invariant,
to contrast the Ricci analysis above.

\begin{figure}[!ht]
\begin{center}
\includegraphics[width=\columnwidth,clip=true]{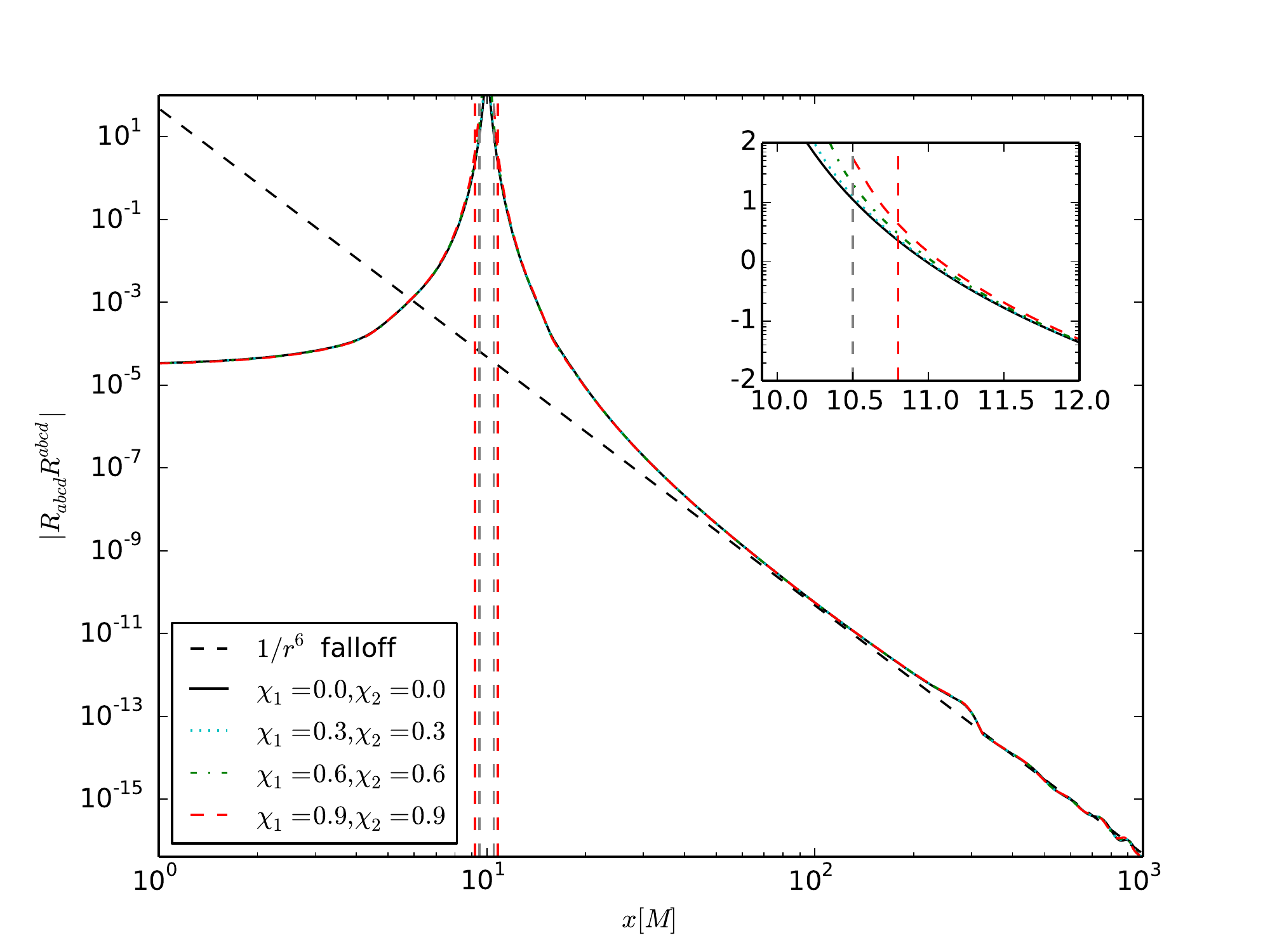}
\end{center}
\caption{
The Kretschmann invariant calculated for the BHB spacetime for
differing values of spin, all of which are aligned with the orbital angular
momentum of the binary, at an initial separation of $r_{12} = 20M$. The $1/r^6$
behavior seen far from the two BHs ($\sim 40M$) is consistent with the
value of the Kretschmann in the single Schwarzschild BH case: $K = 48 M^2/r^6$, where 
$M$ is the total mass of the binary centered on the origin. 
The inset shows the behavior close to the horizon, where the spin effects 
become noticeable.
The Kretschmann becomes large as it approaches a BH, 
because the invariant blows up at a true singularity.}
\label{fig:kretsch}
\end{figure}

As we can see from Fig.~\ref{fig:kretsch}, the Kretschmann invariant
becomes large near the BHs, and falls off as $1/r^6$ as we move
away from the BHs. This means that any error in the FZ will be divided
by a very small number, and so the relative Kretschmann will be large in the
FZ. Therefore, in the weak gravitational field, the relative Kretschmann cannot be used to meaningfully 
measure the accuracy.
On the other hand, it will be extremely small in the IZ where it is being divided
by a very big number. Thus, when the true gravitational field is strong,
the relative Kretschmann can be used as a meaningful measure of the spacetime accuracy.

\begin{figure}[!ht]
\begin{center}
\includegraphics[width=\columnwidth,clip=true]{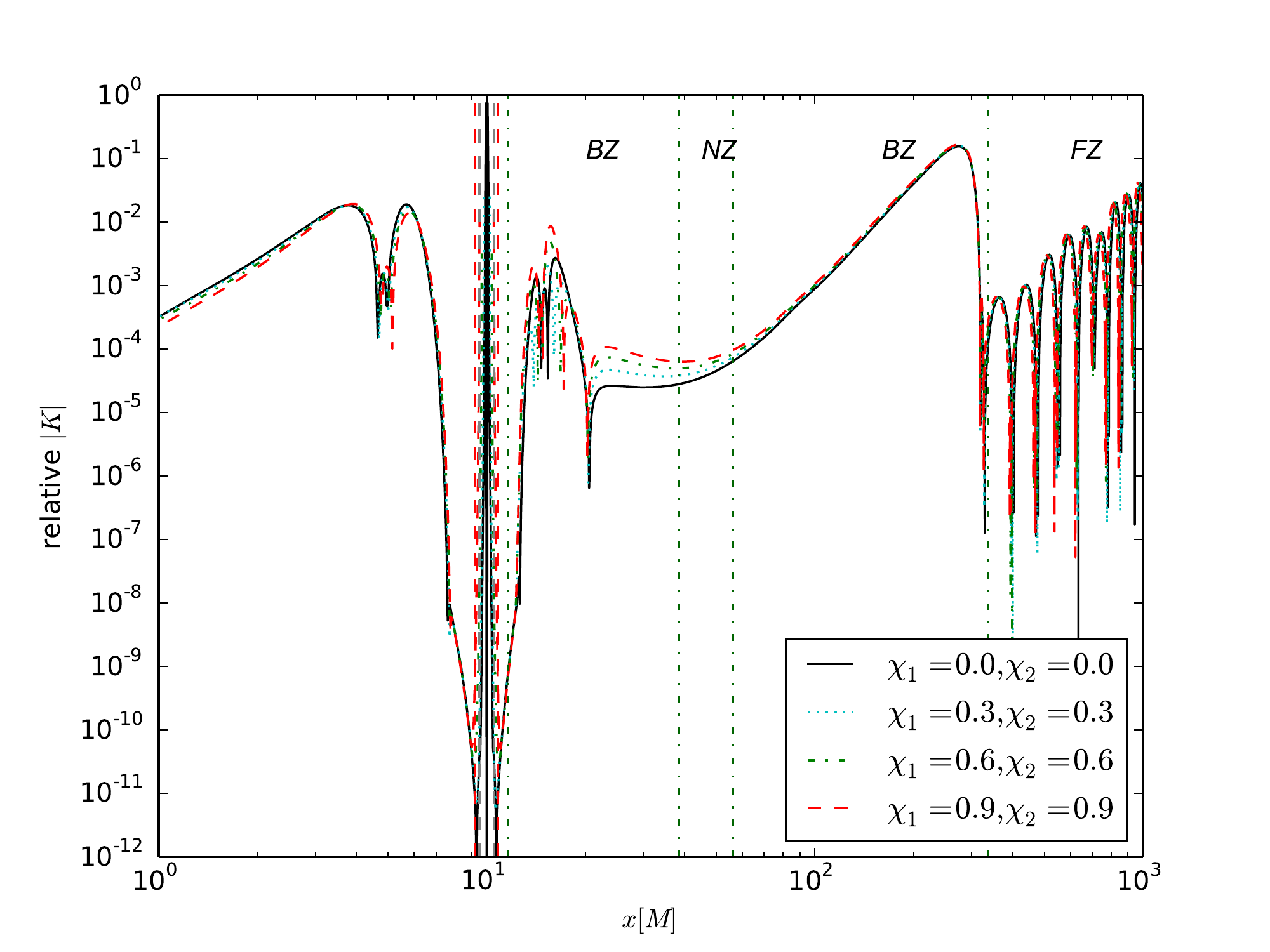}
\end{center}
\caption{The relative Kretschmann for an initial separation of $r_{12} = 20M$,
with a grid resolution of $0.0125$. In this plot, we are plotting aligned
spinning BHs, and increasing the dimensionless spin parameter $\chi_i$ from non
spinning to highly spinning. Observe the normalized behavior that these plots
exhibit. The violation to this invariant is very good close to the horizon due
to the way that we are normalizing. See Fig.~\ref{fig:kretsch} for the
normalization function.}\label{fig:relK}
\end{figure}

\begin{figure}[!ht]
\begin{center}
\includegraphics[width=\columnwidth,clip=true]{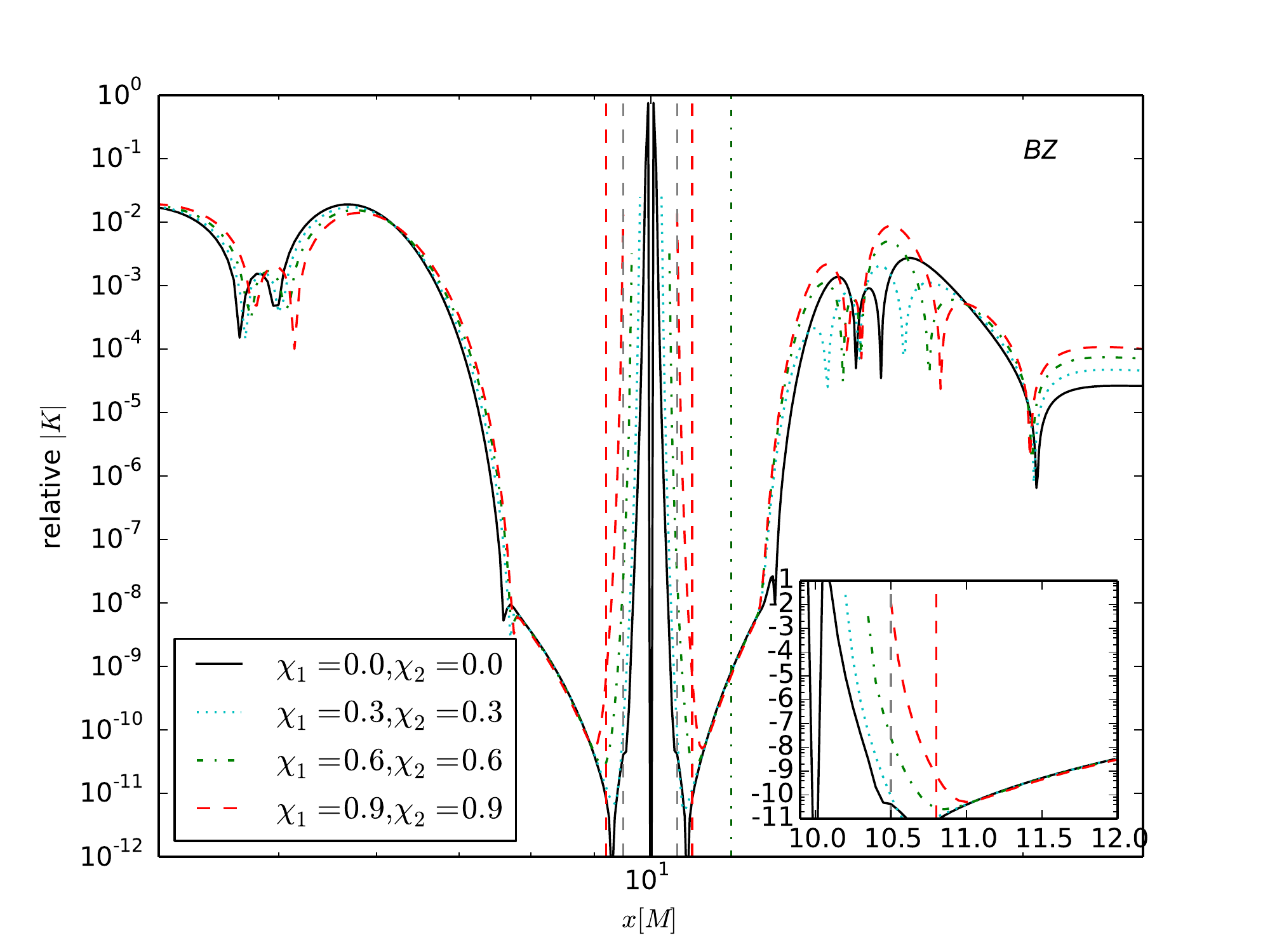}
\end{center}
\caption{Zoomed in view of Fig.~\ref{fig:relK} around the IZ. The inset shows the behavior close to the horizon.}
\label{fig:relKzoom}
\end{figure}

\begin{figure}[!ht]
\begin{center}
\includegraphics[width=\columnwidth,clip=true]{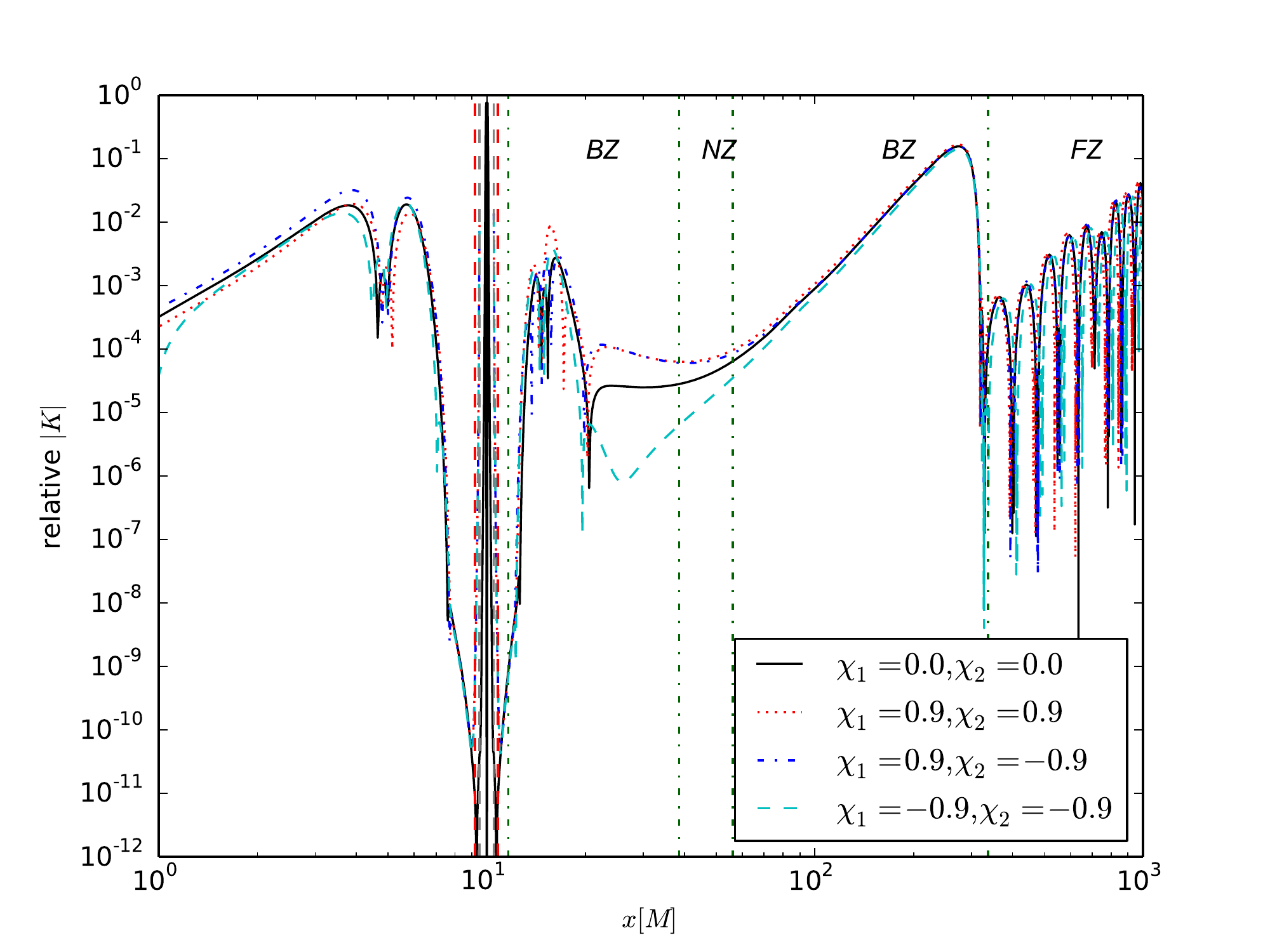}
\end{center}
\caption{The relative Kretschmann for an initial separation of $r_{12} = 20M$,
with a grid resolution of $0.0125$ for varying values of the dimensionless spin
parameter $\chi_i$. In this plot, we are looking at what the relative Kretschmann
does for anti-aligned BHs with high $\chi_i$ values. Observe the normalized
behavior that these plots exhibit. The violation to this invariant is very good
close to the horizon, due to the way that we are normalizing. See
Fig.~\ref{fig:kretsch} for the normalization function.}
\label{fig:relK_anti}
\end{figure}

\begin{figure}[!ht]
\begin{center}
\includegraphics[width=\columnwidth,clip=true]{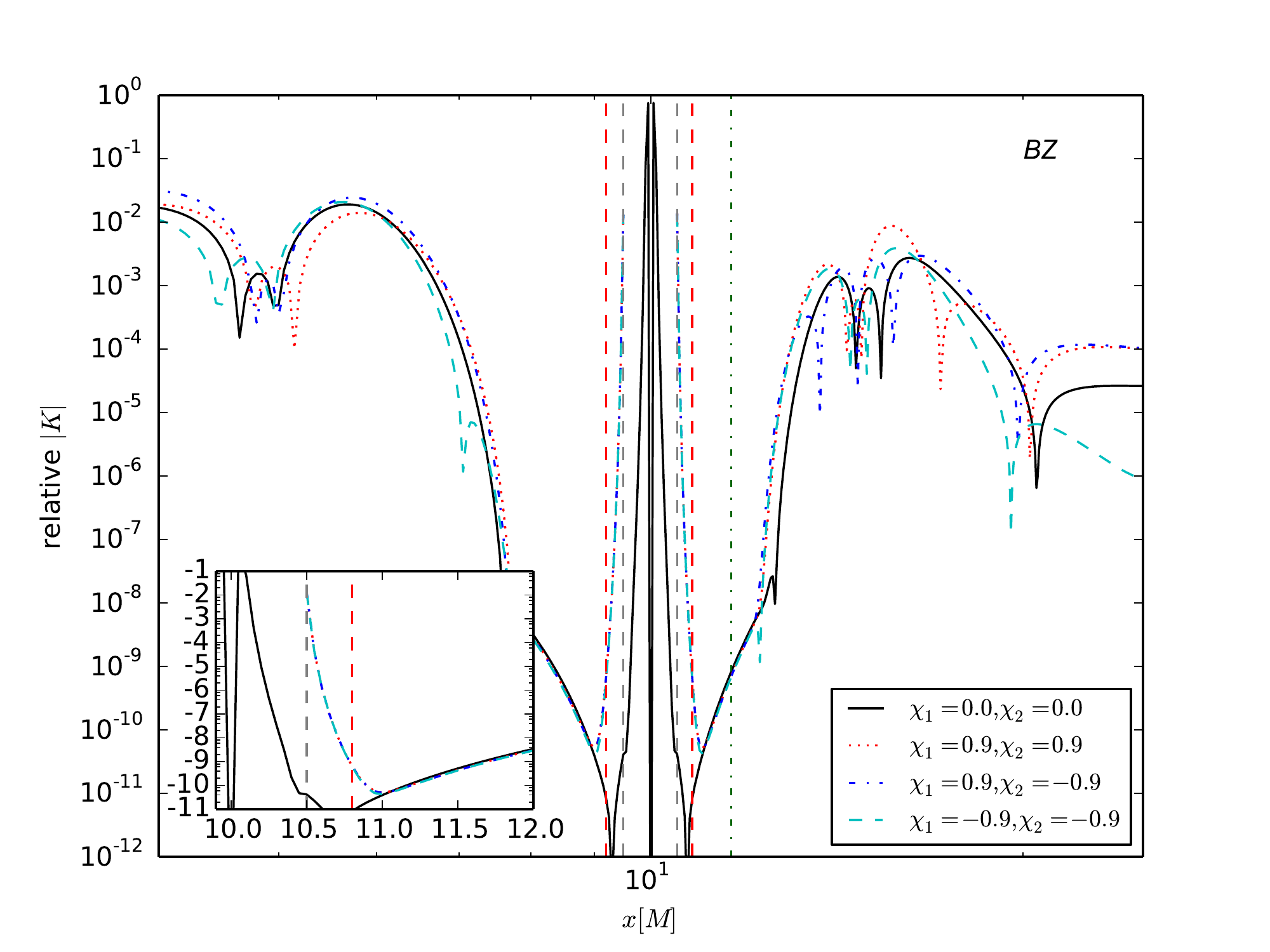}
\end{center}
\caption{Zoomed in view of Fig.~\ref{fig:relK_anti} around the IZ.
The inset shows the behavior close to the horizon.}
\label{fig:relKantizoom}
\end{figure}

We briefly discuss the FZ behavior in Figs.~\ref{fig:relK} and \ref{fig:relK_anti}.
In the FZ, the solution does not follow the $1/r$ behavior that is expected.
This is because the coefficients are calculated in the PN approximation.
We can show this schematically as follows. Imagine a schematic expression of
the FZ metric,
\bea
h_{\mu\nu}^{\rm FZ} \sim \frac{H_{\mu\nu}}{r} \exp(-i \omega (t-r)) + O(1/r^2) \,,
\eea
where $H_{\mu\nu}$ is evaluated from the PN multipole sources. When we plot the
Ricci scalar in the FZ, we have a $1/r$ factor that will act as a damping term.
However, in the relative Kretschmann calculation, this $1/r$ dependence cancels
out due to the $r$ dependence in the Kretschmann invariant. Therefore, any
error accumulated in the PN expression for $H_{\mu\nu}$ will become an
important contribution to the overall error in the FZ, leading to some large
finite amplitude of the Kretschmann at large $r$.

The Kretschmann invariant is not an independent measure of the error in the
approximations, just another way to look at the violation of the spacetime.
The Ricci scalar contains information about 
the spacetime violation, and since the behavior of the Ricci
scalar in the FZ is damped as expected, the observed behavior in the 
relative Kretschmann is not necessarily a concern given the issues with 
this diagnostic discussed earlier due to
the finite PN expression being divided by a small value of the Kretschmann.
This is by no means a proof that the error in the FZ is not dominated by noise
that could be suppressed by taking the code to a higher precision, however,
because the Ricci scalar shows good behavior in the FZ, it is not
worth the analysis that would be required since other complementary
invariants have shown excellent small violations in the FZ.

For these reasons, although the relative Kretschmann is a good way to measure
errors in the IZ, where the fields are strong and dynamical, it is a poor way
to measure the overall accuracy of the global metric in the weak field limit
far from the compact sources.

%%%%%%%%%%%%%%%%%%%%%%%%%%%%%%%%%%%%%%%%
\subsection{Numerical methods}\label{sec:Numerical_methods}
%%%%%%%%%%%%%%%%%%%%%%%%%%%%%%%%%%%%%%%%

\begin{figure}[!ht]
\begin{center}
\includegraphics[width=\columnwidth,clip=true]{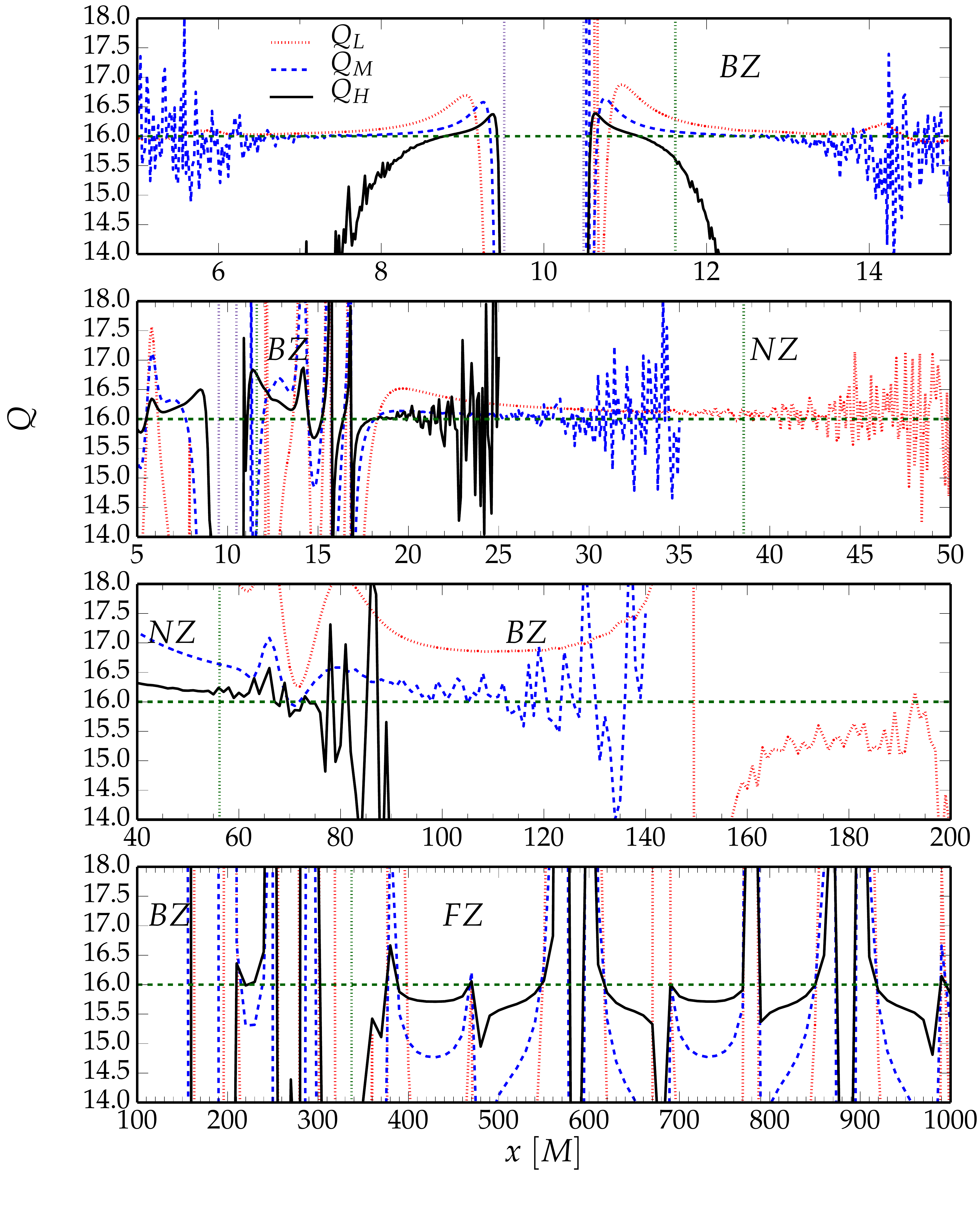}
\end{center}
\caption{
Convergence factor along $x$. Each of the above panels show the Ricci scalar
convergence factor $Q$ at $t=0$ along the coordinate $x$-axis for different
intervals in $x$, depending on the set of mesh spacing used in the convergence
factor computation. On all panels the red dotted line represents the
convergence factor for the low resolution mesh spacing set used in that panel,
while the blue dashed and black solid lines the medium and high resolution mesh
spacing set, respectively.  The green horizontal dashed line at $Q=16.0$ is the
theoretical solution at infinite resolution $h=0$. On the top panel, the low
resolution mesh set, $h_L/M=\{0.1,\,0.05,\,0.025\}$, is used to compute $Q_L$, while
the medium and high resolution ones are $h_M/M=\{0.05,\,0.025,\,0.0125\}$ and
$h_H/M=\{0.025,\,0.0125,\,0.00625\}$, respectively. We masked out $Q_H$ outside the
interval $[7M,\,13M]$ to avoid noise due to round-off precision errors.  On the
second panel from above, the low resolution set is $h_L/M=\{0.8,\,0.4,\,0.2\}$, while
the medium and high ones are $h_M/M=\{0.4,\,0.2,\,0.1\}$ and $h_H/M=\{0.2,\,0.1,\,0.05\}$,
respectively. We also mask out $Q_H$ and $Q_M$ for $x>25M$ and $x>35M$,
respectively, to avoid round-off noise. On the third panel, we use
$h_L/M=\{6.4,\,3.2,\,1.6\}$, $h_M/M=\{3.2,\,1.6,\,0.8\}$ and $h_L/M=\{1.6,\,0.8,\,0.4\}$. We mask
out $Q_H$ and $Q_M$ for $x>90M$ and $x>140M$, respectively. Finally on the bottom
panel, we use $h_L/M=\{51.2,\,25.6,\,12.8\}$, $h_M/M=\{25.6,\,12.8,\,6.4\}$ and
$h_L/M=\{12.8,\,6.4,\,3.2\}$. These different sets of resolutions were used here 
to emphasize the mesh spacings required to be in the convergence regime for
different zones.
}
\label{fig:Q_x}
\end{figure}

In order to compute the several geometric quantities needed for our analysis
partial derivatives of the metric components are needed. One could try to
obtain these derivatives analytically, in a closed form, for each piece of the
metric used when composing the global metric, however this would result into
extremely large expressions which could potentially defeat the goal of
obtaining analytic approximations of BHB spacetimes that are cheaper to compute
than a full general relativistic numerical computation. In addition this fully
analytic approach to the computation of the derivatives would be extremely
tricky to implement in the BZs where not only we need to worry about the metric
matching but also the matching of its derivatives. These analytic complications
give us incentive to compute these derivatives numerically. In all
computations showed in this paper we have discretized the partial derivatives
using a centered, fourth order finite difference stencil.  
In Fig.~\ref{fig:Q_x} we show the Ricci scalar convergence factor
($Q(t=0,x)=(R^{4h}-R^{2h})/(R^{2h}-R^h)=2^p+O(h)$, where $h$ is the mesh
spacing and $p=4$ in our case. See Ref.~\cite{Mundim:2013vca} for more
details) for several different resolutions demonstrating convergence to the
continuum solution to the $4th$ order of approximation.  Since the solution
spans several scales of length, different requirements in terms of mesh
spacings is needed to resolve the solution. For example, on the top panel the
highest resolutions used to compute the convergence factor,
$h_H/M=\{0.025,\,0.0125,\,0.00625\}$, does resolve well the solution features in the
vicinity of the BH location, $x_{{\rm BH},1}/M=10$ in this case.  In addition we
can see clearly the convergence factor tending to $16.0$ around $x/M=9$ or $x/M=11$
as we increase the resolution used in $Q_L$ to the ones in $Q_M$ and $Q_H$, a
clear indication of $4th$ order convergence.  However it is interesting to note
that the high resolution mesh spacing set drives the convergence factor outside
the convergence regime approximately outside the interval $[7M,\,13M]$. This is
mainly due to the limited precision to represent numbers in these computations
(double precision in our case).  Subtractions of very similar numbers results
in catastrophic loss of precision which in turn results in poor convergence
order computation.  A quadruple precision version of the code was used in the
past to evaluate and confirm this loss of precision, however its general use
for our current simulations is prohibitively expensive and we do not report its
results here.

One interesting reading from Fig.~\ref{fig:Q_x} concerns identifying the mesh
spacing requirement for each of the zones describing our metric. For example, in
the vicinity of the BH location, $[10.5M,\,11.5M]$ we can safely say that the set
of mesh spacings $h/M=\{0.00625,\,0.0125,\,0.025,\,0.05\}$
lies within the convergence radius of the $4th$ order scheme.
As this $x$ interval extends beyond, roughly $[11.5M,\,20.0M]$,
the requirement changes to the set of $h/M=\{0.0125,\,0.025,\,0.05,\,0.1\}$.
As we increase the interval farther away, $[20.0M,\,50.0M]$, the set of
$h/M=\{0.1,\,0.2,\,0.4,\,0.8\}$ seem reasonable. As we go farther away the BH 
location the resolution, $[50M,\,200M]$, the resolution requirement drops
for $h=\{0.8,\,1.6,\,3.2,\,6.4\}$ approximately. Finally as we extend to intervals
of $[200M,\,1000M]$ and beyond, mesh spacings of $h/M=\{3.2,\,6.4,\,12.8,\,25.6\}$
seem reasonable to obtain converging solutions. From these studies 
it is clear then that we are able to obtain $4th$ order converging solutions 
from the IZ to the FZ if we are careful in selecting the appropriate 
mesh spacings. 

%%%%%%%%%%%%%%%%%%%%%%%%%%%%%%%%%%%%%%%%
\subsection{Orbital hang-up effect, and long time evolutions of the BHB}\label{sec:hangup}
%%%%%%%%%%%%%%%%%%%%%%%%%%%%%%%%%%%%%%%%

In Ref.~\cite{Gallouin:2012kb}
the Ricci violation was shown for an initial spatial hypersurface.
This work has presented figures for the Ricci violation
and relative Kretschmann by using Eq.~(\ref{eq:CTforLTE})
as the coordinate transformation for long time evolutions 
of our dynamic spacetime,
these figures just capture a snapshot at an instant in time
which is basically the same procedure that was used for the initial data.
This means that we just need the EOM at an instant in time,
and it is not necessary to solve the EOM.
Therefore, it is not clear that we have introduced an appropriate EOM
for long time evolutions of the BHB system,
and thus to confirm our results 
we present the orbital evolutions.

A natural way to test this implementation of the EOM
is to see if this work can reproduce any of the known effects of spin dynamics 
in aligned non-precessing systems, such as the orbital hang-up effect discovered 
in Ref.~\cite{Campanelli:2006uy}. 
Recovering the hang up effect is an easy way to show the correctness in the implementation
of the EOM for the binary. 
The orbital hang-up effect is an effect where the spin of the individual BHs 
add to the orbital angular momentum of the binary, causing the orbit to inspiral more slowly, 
as it has to dissipate more angular momentum. This leads to a pile up of the orbits, 
causing the ``hang-up". For the following plot, we will be considering equal mass BHs 
in quasi circular orbits, with dimensions in terms of the total mass $M = m_1 + m_2$ of the binary.
This effect was shown to be the strongest at
merger in Ref.~\cite{Campanelli:2006uy}, but as we show here it also has an effect 
in the PN regime.

\begin{figure}[ht!]
\begin{center}
\includegraphics[width=\columnwidth,clip=true]{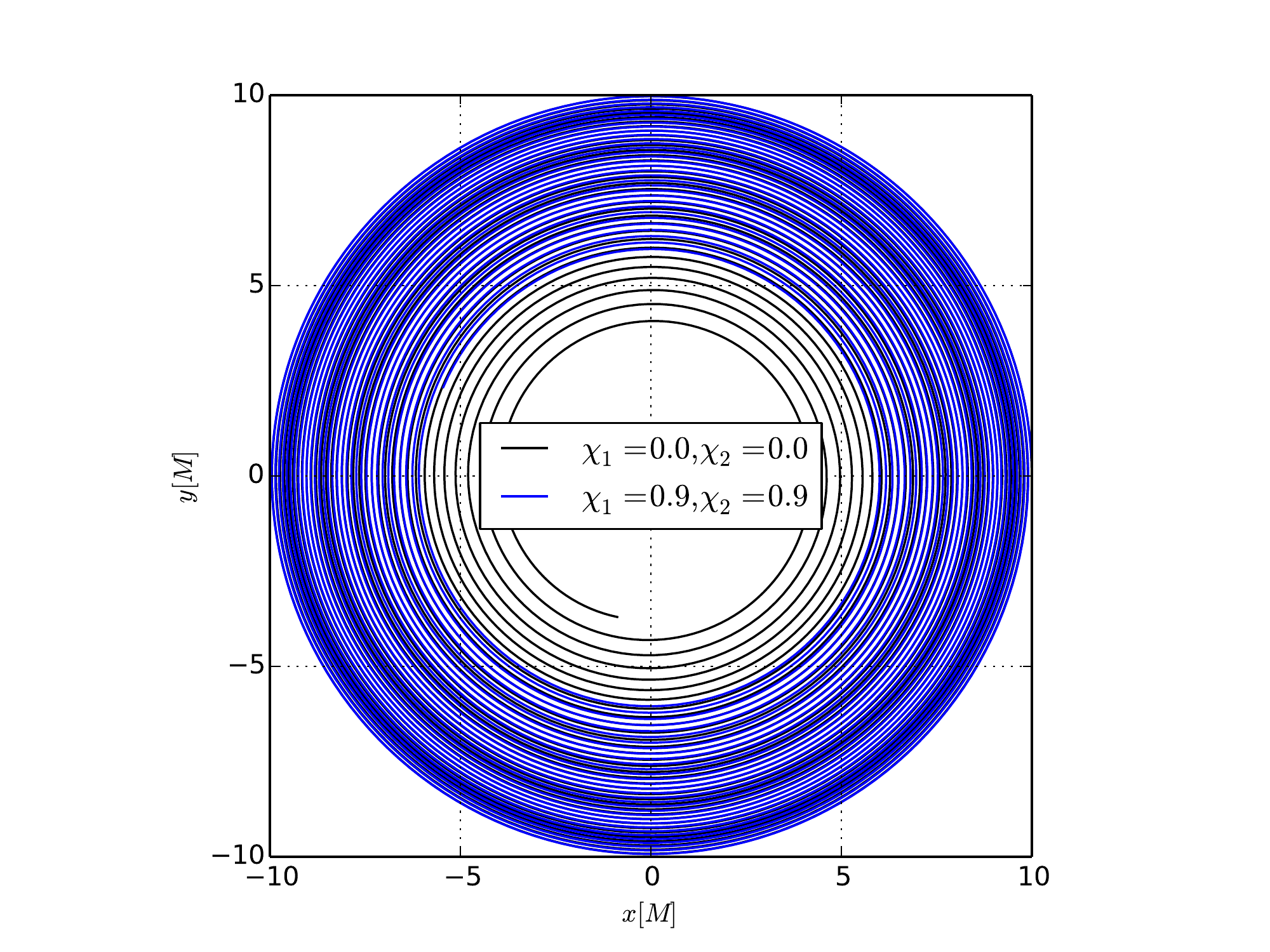}
\includegraphics[width=\columnwidth,clip=true]{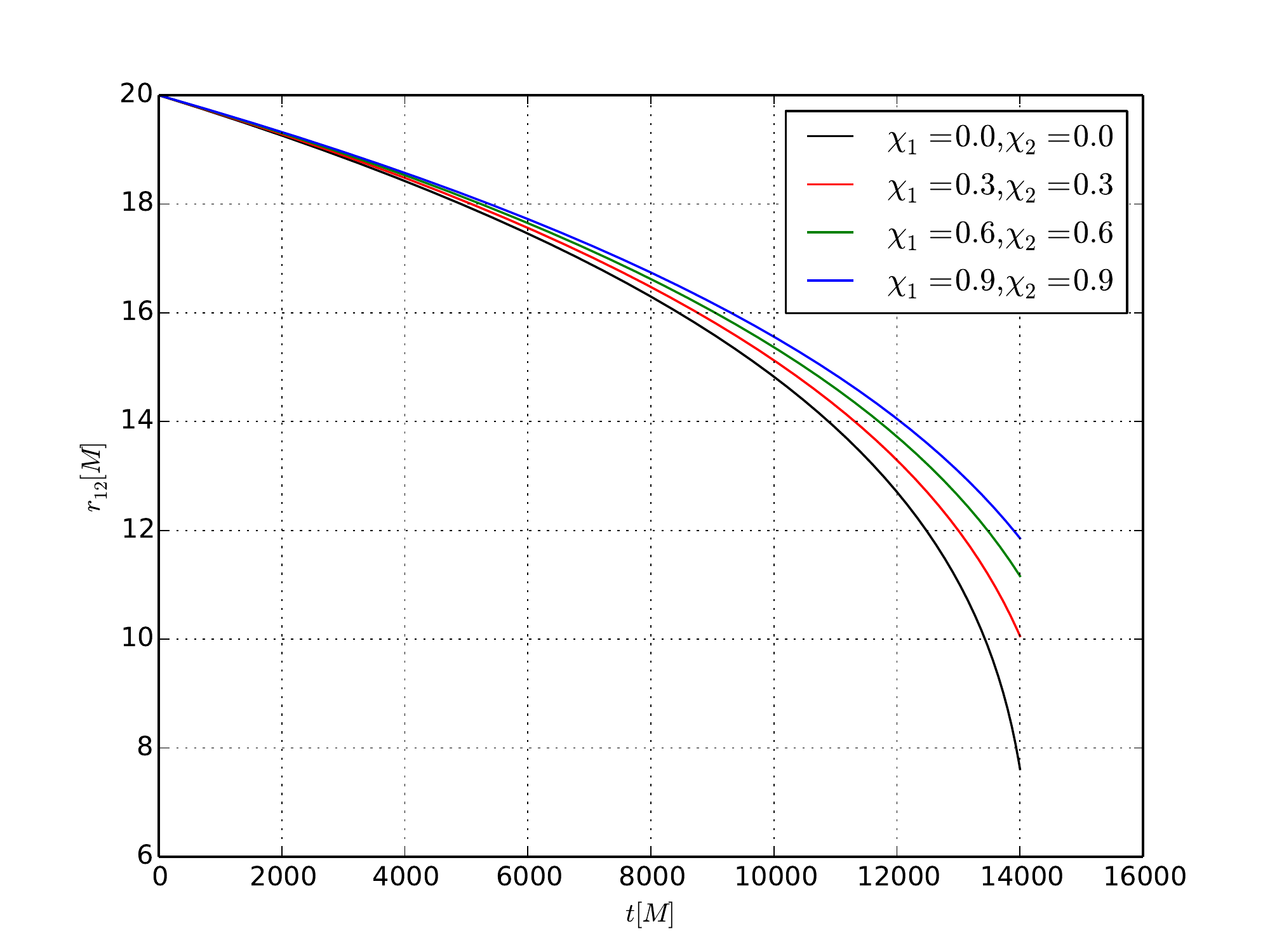}
\end{center}
\caption{The top panel is the orbital hang-up effect shown by plotting the individual trajectories of BH1.
Note that the orbits get bunched up as the spin is increased from $\chi_1=\chi_2=0.0$ to $0.9$.
Both of the BH spins are aligned with the orbital angular momentum of the binary.
The bottom panel shows the orbital hang-up effect as a decrease in the orbital separation as
a function of time for varying aligned spin parameters, with the fiducial spin values chosen to 
coincide with the spins used in the Ricci and relative Kretschmann analyzes. 
For both the top and bottom panels, the evolution was terminated at $14000M$ in time. 
}
\label{fig:hangup_traj}
\end{figure}

This effect can be seen clearly in Fig.~\ref{fig:hangup_traj}. Had the spins been anti-aligned, 
the reverse would be seen, with the highly spinning BHs plunging quickly compared 
to the non spinning case. Comparisons to the orbital dynamics in Refs.~\cite{Noble:2012xz} 
and \cite{Scheel:2015CQG} yield good confirmation between the trajectory plotted 
and the non spinning trajectory here, offering reassurance that the correct dynamics 
are being calculated in the PN approximation, and thus that our results are valid for long time evolutions.

%%%%%%%%%%%%%%%%%%%%%%%%%%%%%%%%%%%%%%%%
\section{DISCUSSION}\label{sec:discussion}
%%%%%%%%%%%%%%%%%%%%%%%%%%%%%%%%%%%%%%%%

We have constructed a globally analytic, approximate BHB spacetime
via asymptotically matching BH perturbation theory to PN formalism farther out
to a PM spacetime even further away. The procedure of asymptotic matching had
to be generalized from Ref.~\cite{Gallouin:2012kb} to be valid on all spatial
hypersurfaces, instead of a small group of initial hypersurfaces near $t -
t_0$. Matching the metrics in this way allows us to construct a global metric,
which is correct until the PN approximation breaks down, around $10M$ in
orbital separation. 

The validity of this global metric was extensively tested using several
different techniques. We first calculated the absolute value of the Ricci
scalar and plotted it along a particular axis and compared it with the exact
solution value $R=0$. Then the evolution of the Ricci scalar was explored, to
ensure that there were no sporadic errors in the evolution that would
contribute unduly to the overall error. Fortunately, the behavior observed in
this evolution was a smooth increase in the violation to the Einstein field
equations due to the slow motion assumption breaking down as the BHs inspiral.

To contrast with the Ricci scalar analysis, we performed an analysis
by using the relative Kretschmann $K_{\rm rel}$ in Eq.~\eqref{eq:relK}.
However, it was noticed that this measure of the
error in the global metric is not the best in the FZ, since the FZ behavior is not
damped, and indeed appears to be the largest contributor of error.  As a result of this
analysis, though the relative Kretschmann has the attractive feature of being an
invariant with a natural scale of comparison, we will be using the Ricci scalar as 
the measure of the accuracy of the global metric out to the FZ, electing to 
use the Kretschmann for studies of the violation close to either BH. 

This metric is valid dynamically and for long time evolutions, which makes it
an ideal metric to use when studying effects happening in the relativistic
regime of BHBs. The immediate application of this metric is to implement it
into the \harm code, which can then be used to study MHD in the context of the
BHB problem with spins. This will allow new studies of accretion physics,
giving us the theoretical tools to make predictions for the EM 
signatures of BHBs with spin. 

Certain practical considerations are required when talking about this global
metric construction. In this paper, we have considered only one choice of
transition function, which is summarized in Appendix~\ref{app:trans_func}. The
transition functions' arguments were chosen by pragmatic considerations;
experimentally shown to give a lower violation to the Einstein field equations,
and not necessarily by any mathematical arguments, such as the Frankenstein
theorems of Ref.~\cite{Yunes:2006mx}.  

A study which will be important for the future is an in depth analysis of
the transition region near the IZ-NZ BZ. From preliminary hydrodynamic
simulations, the mini disks around the individual BHs fall entirely in the BZ,
and the choice of transition function may have a significant
impact on the spacetime, and thus the gas dynamics will be impacted. Practical
choices of the free parameters need to be explored further and will be highly
valued for upcoming MHD runs.

Another test that can be done is to explore test particle trajectories in the
spacetime. This powerful tool will give us a good idea of how the spacetime is
doing complementary to the violation of the Ricci scalar and relative
Kretschmann invariant. This is being developed currently, and will be explored
in detail in a future paper.

A final step for this project is to generalize this to arbitrarily aligned spins, which will
lead to precession. This has the added benefit of being able to study even more
interesting spin effects such as transitional precession and spin flips. This
will require refinement in our techniques. The IZ metric will have to take into
account that the tidal fields are no longer solely along the $r$ direction,
but shifted so that the spin axis is arbitrarily aligned with the orbital
angular momentum. The EOM 
need to be updated to take into
account the higher order spin dynamics, and the metric will need to be modified
to generate spins along any direction. Though this process will be arduous, a
fully analytic spacetime describing a precessing, arbitrarily aligned,
spinning BHB spacetime will allow for GRMHD simulations to explore
completely new territory of gas dynamics in the context of precessing BH spins.

%%%%%%%%%%%%%%%%%%%%%%%%%%%%%%%%%%%%%%%%
\acknowledgments

We thank Carlos O. Lousto, Yosef Zlochower, Dennis Bowen and
Nicol\'{a}s Yunes for a careful reading of this
manuscript, and for helpful discussions.  B.~I. and M.~C. are supported by NSF
grants  AST-1516150, AST-1028087 and PHY-1305730.  
B.~C.~M.~ is supported by the LOEWE-Program in HIC for FAIR.
H.~N. acknowledges support by 
MEXT Grant-in-Aid for Scientific Research
on Innovative Areas,
``New Developments in Astrophysics Through Multi-Messenger Observations
of Gravitational Wave Sources'', No.~24103006. Authors also
gratefully acknowledge the NSF for financial support from Grant
PHY-1305730. Computational resources were provided by XSEDE allocation
TG-PHY060027N, and by the BlueSky Cluster 
at Rochester Institute of Technology, which were supported
by NSF grant No. AST-1028087, and PHY-1229173.

%%%%%%%%%%%%%%%%%%%%%%%%%%%%%%%%%%%%%%%%
\appendix

%%%%%%%%%%%%%%%%%%%%%%%%%%%%%%%%%%%%%%%%
\section{Transition functions}\label{app:trans_func}
%%%%%%%%%%%%%%%%%%%%%%%%%%%%%%%%%%%%%%%%

When constructing the global metric, we must introduce appropriate transition
functions in the BZs to avoid erroneous behaviors~\cite{Yunes:2006mx}.
In this analysis, we follow Ref.~\cite{Mundim:2013vca}, and use the following
transition function:
\begin{widetext}
\begin{align}
\label{trans-eq}
f(r,\,r_0,\,w,\,q,\,s) =
\left\{
	\begin{array}{lr}
        0 \,, & \hspace{-85mm} r \le r_0 \,, \\ 
        \displaystyle{\frac{1}{2}} \left\{ 1 +\tanh \left[\frac{s}{\pi} \left(\chi(r,\,r_0,\,w) - \frac{q^2}{\chi(r,\,r_0,\,w)} \right) \right] \right\} \,, 
            & \hspace{9mm}r_0 < r < r_0 + w \,, \\
        1 \,, & \hspace{-85mm} r \ge r_0 + w \,,     
    	\end{array}
    \right.
\end{align}
\end{widetext}
where $\chi(r, r_0, w)=\tan[\pi(r - r_0)/(2w)]$, and $r_0$, $w$, $q$ and $s$
are parameters.  Great detail on this transition function can be found in
Refs.~\cite{JohnsonMcDaniel:2009dq,Yunes:2005nn,Yunes:2006iw}.  This transition
function uses different parameters in each of the BZs, that are
modified from Ref.~\cite{JohnsonMcDaniel:2009dq}.

In the analysis, we started by using the parameters from Ref.~\cite{Mundim:2013vca}:
\begin{align}
f_{\rm near} = f(x, \, 2.2m_2 - m_1 r_{12}/M, \, r_{12} - 2.2M, \, 1, \, 1.4), \\
f_{\rm inner, \, A} = f(r_A, \, 0.256r^T_A, \, 3.17(M^2 r^5_{12})^{1/7}, \, 0.2, \, r_{12}/M).
\end{align}
Here $r^T_A$ is the transition function radius, derived by requiring the
uncontrolled remainders of the IZ and NZ approximations be roughly equal.  The
NZ-FZ transition function is unchanged with respect to
Ref.~\cite{Mundim:2013vca}, and the details of this choice of transition
function can be explored in that paper.  It should be noted that although we
should formally use the transition functions given in
Ref.~\cite{Gallouin:2012kb} due to the matching order according to the
Frankenstein theorems of Ref.~\cite{Yunes:2006mx}, we have used the above
transition functions because it was found by experiment that they give overall
better results in the Ricci calculation.  While this is not mathematically
rigorous, it is a practical choice that we made to minimize the violation to
the Einstein field equations represented by the Ricci scalar and the relative
Kretschmann invariant. 

It is also noted here that in practice we use the value $s = b/M$ where $b$ is
a (constant) initial orbital separation, as opposed to $r_{12}/M$ which is time
dependent.  This choice was made in Ref.~\cite{Mundim:2013vca}, so to compare
as directly as possible, we use the same $s$ parameter as is used there.

In the course of doing large separation runs ($\approx 100M$) with the non
spinning global metric, a problem was found with the IZ-NZ transition function.
At large separation, the value of $s = r_{12}/M$ becomes large.  As discussed
in Ref.~\cite{Yunes:2006iw}, $q$ determines the location where the transition
function equals $1/2$, and $s$ sets the slope, given by $s(1 + q^2)/(2 w)$.
When we discuss the Ricci scalar of the spacetime,  $s$ becomes more sensitive
than $q$ in setting the slope for fixed $r_0$ and $w$.  Due to the value of
$r_{12}$ setting the slope, in large separation runs with $m_1 = m_2 = m/2$,
the second derivative of the transition function $f_{\rm inner, \, A}$ can
become very large.  Because of this, a new value of $s$ is suggested to
minimize the absolute value of the second derivative of this transition
function.  Due to analytic testing of the transition function, this $s$
parameter was set to $12$ for our future work and implementation into 
\harm.

Of course, there are many other parameters that set the transition function,
all of which can have wildly different values.  In future work, it
would be nice to have a way of optimizing the parameters to give the minimum
violation to the Ricci scalar.  This could be achieved by using a Monte Carlo
simulation to explore this parameter space and pinpoint the ideal values of
the different parameters, while still being a valid transition function 
(i.e. obeying the Frankenstein theorems~\cite{Yunes:2006mx}). This will be reserved
for future work.

%%%%%%%%%%%%%%%%%%%%%%%%%%%%%%%%%%%%%%%%
\section{Ingoing Kerr Coordinates to Cook-Scheel Harmonic Coordinates}\label{app:IK_to_CSH}
%%%%%%%%%%%%%%%%%%%%%%%%%%%%%%%%%%%%%%%%

The NZ metric is calculated in the PN harmonic coordinates,
but to accurately describe gas dynamics close to the event horizons, 
it is desirable to have the horizon penetrating property.
Therefore, the Cook-Scheel harmonic
(CS-H) coordinates~\cite{Cook:1997qc} are ideal
for the Kerr spacetime, which describes the background IZ metric.

Here we change the notation used throughout the paper
slightly because this coordinate transformation 
is for a single BH. Therefore, in this appendix \emph{only}, 
we will take $M$ to be the mass of the individual Kerr BH, 
and not the total mass.

The IZ metric presented in Ref.~\cite{Yunes:2005ve}, however, is in the ingoing
Kerr (IK) coordinates.  Noting the similarities between the more familiar
Boyer-Lindquist (BL) coordinates, $r_{\rm IK} = r_{\rm BL}$ and $\theta_{\rm
IK} = \theta_{\rm BL}$, we can rewrite the coordinate transformation from the
IK ($v_{\rm IK},\,r_{\rm IK},\,\theta_{\rm IK},\phi_{\rm IK}$) coordinates to
CS-H ($t_{\rm H},\,x_{\rm H},\,y_{\rm H},\,z_{\rm H}$) coordinates~\footnote{
Here, the changed notation of the CS-H coordinates is one of convenience.  We
add the subscript H so as not to confuse ourselves.}:
\begin{align}
t_{\rm H} &= v_{\rm IK} - r_{\rm IK}
+ 2 \,M \,\ln \left|\frac{2\,M}{r_{\rm IK}-r_{-}}\right|
\,,
\nonumber \\ 
x_{\rm H} + i\,y_{\rm H} &= (r_{\rm IK}-M+i\,a) e^{i\,\phi_{\rm IK}} \,\sin \theta_{\rm IK} \,,
\nonumber \\ 
z_{\rm H} &= (r_{\rm IK} - M) \cos \theta_{\rm IK} \,.
\end{align}
It is noted that the following calculations are similar to to the summary in
the appendix of Ref.~\cite{Yunes:2005ve} for the coordinate transformation
between the IK and Kerr-Schild coordinates.  To calculate the Jacobian to
transform tensors, we rewrite the above relations as
\begin{align}
x_{\rm H} &= \left[ (r_{\rm IK}-M) \,\cos \phi_{\rm IK}
 - a \,\sin \phi_{\rm IK} \right] \sin \theta_{\rm IK} \,,
\nonumber \\
y_{\rm H} &= \left[ (r_{\rm IK}-M) \,\sin \phi_{\rm IK}
 + a \,\cos \phi_{\rm IK} \right] \sin \theta_{\rm IK} \,,
\nonumber \\
z_{\rm H} &= (r_{\rm IK} - M) \cos \theta_{\rm IK} \,.
\end{align}
Here, we calculate the radial coordinate in the CS-H as
\begin{align}
r_{\rm H}^2 &= x_{\rm H}^2 + y_{\rm H}^2 + z_{\rm H}^2
\cr 
&= r_{\rm IK}^2 - 2 M r_{\rm IK} + M^2+a^2-a^2 \cos^2 \theta_{\rm IK} \,.
\end{align}
When we solve the above equation with respect to $r_{\rm IK}$,
there are four solutions, and one of the solutions,
\begin{align}
r_{\rm IK} =&
\frac{1}{2} 
\biggl[ 2 \sqrt{r_{\rm H}^4-2 a^2 r_{\rm H}^2+a^4+4 a^2 z_{\rm H}^2}
\cr & +2 r_{\rm H}^2-2 a^2 \biggr]^{1/2} + M \,,
\end{align}
gives the appropriate radial coordinate for large $r_{\rm H}$.
Therefore, the inverse transformation is summarized as
\begin{align}
r_{\rm IK} &= \sqrt{\frac{r_{\rm H}^2-a^2+W}{2}} + M \,,
\nonumber \\
\theta_{\rm IK} &= \arccos \frac{z_{\rm H}}{(r_{\rm IK} - M)} \,,
\nonumber \\ 
\phi_{\rm IK} &= \arctan {\frac {(r_{\rm IK} - M)\,y_{\rm H} - a\,x_{\rm H}}
{(r_{\rm IK} - M)\,x_{\rm H} + a\,y_{\rm H}}} \,,
\end{align}
and 
\begin{align}
v_{\rm IK} &= t_{\rm H} + r_{\rm IK}
- 2 \,M \,\ln \left|\frac{2\,M}{r_{\rm IK}-r_{-}}\right|
\,,
\end{align}
where 
\begin{align}
W &= \sqrt{(r_{\rm H}^2-a^2)^2 + 4\,a^2\,z_{\rm H}^2} \,.
\end{align}
We have the useful relations,
\begin{align}
\sin \theta_{\rm IK} &= \sqrt{\frac{x_{\rm H}^2+y_{\rm H}^2}{(r_{\rm IK}-M)^2+a^2}} \,,
\nonumber \\ 
\sin \phi_{\rm IK}  &= 
\frac{(r_{\rm IK} - M)\,y_{\rm H} - a\,x_{\rm H}}
{[((r_{\rm IK}-M)^2+a^2)(x_{\rm H}^2+y_{\rm H}^2)]^{1/2}} \,,
\nonumber \\
\cos \phi_{\rm IK} &= 
\frac{(r_{\rm IK} - M)\,x_{\rm H} + a\,y_{\rm H}}
{[((r_{\rm IK}-M)^2+a^2)(x_{\rm H}^2+y_{\rm H}^2)]^{1/2}} \,.
\end{align}

Using the above inverse transformation,
the Jacobian for this coordinate transformation, 
$\partial x^a_{\rm IK}/\partial x^b_{\rm H}$ is calculated as
\begin{align}
\frac{\partial v_{\rm IK}}{\partial t_{\rm H}}
=& 1 \,,
\nonumber \\ 
\frac{\partial v_{\rm IK}}{\partial x_{\rm H}}
=& \frac{x_{\rm H}}{2\,(r_{\rm IK}-M)}
\left(1 + \frac{r_{\rm H}^2-a^2}{W} \right) \left(1+\frac{2\,M}{r_{\rm IK}-r_-} \right) \,,
\nonumber \\ 
\frac{\partial v_{\rm IK}}{\partial y_{\rm H}}
=& \frac{y_{\rm H}}{2\,(r_{\rm IK}-M)}
\left(1 + \frac{r_{\rm H}^2-a^2}{W} \right) \left(1+\frac{2\,M}{r_{\rm IK}-r_-} \right) \,,
\nonumber \\ 
\frac{\partial v_{\rm IK}}{\partial z_{\rm H}}
=& \frac{z_{\rm H}}{2\,(r_{\rm IK}-M)}
\left(1 + \frac{r_{\rm H}^2+a^2}{W} \right) \left(1+\frac{2\,M}{r_{\rm IK}-r_-} \right) \,,
\nonumber \\ 
\frac{\partial r_{\rm IK}}{\partial t_{\rm H}}
=& 0 \,,
\nonumber \\ 
\frac{\partial r_{\rm IK}}{\partial x_{\rm H}}
=& \frac{x_{\rm H}}{2\,(r_{\rm IK}-M)}
\left(1 + \frac{r_{\rm H}^2-a^2}{W} \right)
\,,
\nonumber \\ 
\frac{\partial r_{\rm IK}}{\partial y_{\rm H}}
=& \frac{y_{\rm H}}{2\,(r_{\rm IK}-M)}
\left(1 + \frac{r_{\rm H}^2-a^2}{W} \right)
\,,
\nonumber \\ 
\frac{\partial r_{\rm IK}}{\partial z_{\rm H}}
=& \frac{z_{\rm H}}{2\,(r_{\rm IK}-M)}
\left(1 + \frac{r_{\rm H}^2+a^2}{W} \right)
\,,
\nonumber \\ 
\frac{\partial \theta_{\rm IK}}{\partial t_{\rm H}}
=& 0 \,,
\nonumber \\ 
\frac{\partial \theta_{\rm IK}}{\partial x_{\rm H}}
=& \frac{x_{\rm H}\,z_{\rm H}}{2\,(r_{\rm IK}-M)^2}
\left(1 + \frac{r_{\rm H}^2-a^2}{W} \right) 
\cr & \times
\left( (r_{\rm IK}-M)^2 - z_{\rm H}^2 \right)^{-1/2}
\,,
\nonumber \\ 
\frac{\partial \theta_{\rm IK}}{\partial y_{\rm H}}
=& \frac{y_{\rm H}\,z_{\rm H}}{2\,(r_{\rm IK}-M)^2}
\left(1 + \frac{r_{\rm H}^2-a^2}{W} \right)
\cr & \times 
\left( (r_{\rm IK}-M)^2 - z_{\rm H}^2 \right)^{-1/2}
\,,
\nonumber \\ 
\frac{\partial \theta_{\rm IK}}{\partial z_{\rm H}}
=& - \left[ 1 - \frac{z_{\rm H}^2}{2\,(r_{\rm IK}-M)^2}
\left(1 + \frac{r_{\rm H}^2+a^2}{W} \right)
\right] 
\cr & \times
\left( (r_{\rm IK}-M)^2 - z_{\rm H}^2 \right)^{-1/2}
\,,
\nonumber \\ 
\frac{\partial \phi_{\rm IK}}{\partial t_{\rm H}}
=& 0 \,,
\nonumber \\ 
\frac{\partial \phi_{\rm IK}}{\partial x_{\rm H}}
=& - \frac{y_{\rm H}}{x_{\rm H}^2+y_{\rm H}^2}
+ \frac{a\,x_{\rm H}}{2\,(r_{\rm IK}-M)((r_{\rm IK}-M)^2+a^2)}
\cr & \times
\left(1 + \frac{r_{\rm H}^2-a^2}{W} \right)
\,,
\nonumber \\ 
\frac{\partial \phi_{\rm IK}}{\partial y_{\rm H}}
=& \frac{x_{\rm H}}{x_{\rm H}^2+y_{\rm H}^2}
+ \frac{a\,y_{\rm H}}{2\,(r_{\rm IK}-M)((r_{\rm IK}-M)^2+a^2)}
\cr & \times
\left(1 + \frac{r_{\rm H}^2-a^2}{W} \right)
\,,
\\ 
\frac{\partial \phi_{\rm IK}}{\partial z_{\rm H}}
=& \frac{a\,z_{\rm H}}{2\,(r_{\rm IK}-M)((r_{\rm IK}-M)^2+a^2)}
\cr & \times
\left(1 + \frac{r_{\rm H}^2+a^2}{W} \right)
\,.
\end{align}
The right hand side of the above equations includes the IK and CS-H coordinates
because the expressions give a compact form.
Although there is a apparent divergent behavior at $r_{\rm IK}=M$,
this can be removed by using
\begin{align}
r_{\rm IK}-M =& \sqrt{\frac{r_{\rm H}^2-a^2+W}{2}} \,.
\end{align}

Finally, the perturbed metric in the IK coordinates is transformed to the CS-H
coordinates as
\bea
g_{\mu\nu}^{\rm H} = \frac{\partial x^{\mu'}_{\rm IK}}{\partial x^\mu_{\rm H}}
\frac{\partial x^{\nu'}_{\rm IK}}{\partial x^\nu_{\rm H}} g_{\mu'\nu'}^{\rm IK} \,.
\eea
This IZ metric in the CS-H coordinates will then be matched to the NZ metric. 

%%%%%%%%%%%%%%%%%%%%%%%%%%%%%%%%%%%%%%%%
\section{Details about the horizon and the innermost stable circular orbit}\label{app:ISCO}
%%%%%%%%%%%%%%%%%%%%%%%%%%%%%%%%%%%%%%%%

In our analysis of the validity of the spacetime,
it is helpful to understand the location of the BH horizon
and the ISCO in the PNH coordinates.
We discuss the coordinate transformation from the BL 
to the CS-H coordinates again, because various useful results have been
derived in the BL coordinates.

The coordinate transformation from the BL coordinates 
($t_{\rm BL},\,r_{\rm BL},\,\theta_{\rm BL},\,\phi_{\rm BL}$)
to the CS-H coordinates ($t_{\rm H},\,x_{\rm H},\,y_{\rm H},\,z_{\rm H}$) is given by
\begin{align}
t_{\rm H} &= t_{\rm BL} + \frac{r_{+}^2+a^2}{r_{+}-r_{-}} 
\,\ln \left|\frac{r_{\rm BL}-r_{+}}{r_{\rm BL}-r_{-}}\right|
\,,
\nonumber \\
x_{\rm H} + i\,y_{\rm H} &= (r_{\rm BL}-M+i\,a) e^{i\,\varphi} \,\sin \theta_{\rm BL} \,,
\nonumber \\
\qquad \varphi &= \phi_{\rm BL} + \frac{a}{r_{+}-r_{-}} 
\,\ln \left|\frac{r_{\rm BL}-r_{+}}{r_{\rm BL}-r_{-}}\right|
\,,
\nonumber \\ 
z_{\rm H} &= (r_{\rm BL} - M) \cos \theta_{\rm BL} \,,
\label{eq:CT}
\end{align}
where $r_{\pm} = M \pm \sqrt{M^2-a^2}$
denote the event horizon ($r_+$) and Cauchy horizon ($r_-$)
in the BL coordinates, and $\varphi$ is same as $\phi_{\rm IK}$ in Appendix~\ref{app:IK_to_CSH}.

The following equations are useful to understand the CS-H coordinates.
\begin{align}
x_{\rm H}^2+y_{\rm H}^2 &= [(r_{\rm BL}-M)^2 + a^2] \,\sin^2 \theta_{\rm BL}
\cr
& = [(r_{\rm BL}-M)^2 + a^2] \left(1 - \frac{z_{\rm H}^2}{(r_{\rm BL}-M)^2} \right)
\,,
\cr
r_{\rm H}^2 &=  (r_{\rm BL}-M)^2 + a^2 \,\sin^2 \theta_{\rm BL}
\,,
\cr 
r_{\rm H}\,\cos \theta_{\rm H}  &= z_{\rm H}  
= (r_{\rm BL} - M) \cos \theta_{\rm BL} \,,
\cr
\phi_{\rm H} &= \arctan \frac{y_{\rm H}}{x_{\rm H}} 
=
\phi_{\rm a} + \varphi \,;
\cr
\phi_{\rm a} &= \arctan \frac{a}{r_{\rm BL}-M} \,.
\end{align}

The event horizon ($r_{\rm BL}=r_+$)
is located at $r_{\rm H}=\sqrt{M^2-a^2\cos^2 \theta_{\rm BL}}$
in the CS-H coordinates from the transformations above.
We also will use this location in the PNH coordinates
as a rough estimation of the event horizon
because the transformation from the CS-H
to the PNH coordinates is treated perturbatively, so the location of the horizon will not change much.
On the equatorial plane ($\theta_{\rm BL} = \theta_{\rm H}  = \pi/2$), we have
the event horizon at
\begin{align}
r_{\rm H} = M \,,
\end{align}
which is independent of the spin parameter, $a$.
It is noted that there is a coordinate singularity
at $r_{\rm BL}=M$, i.e., $r_{\rm H}=|a| \sin \theta_{\rm BL}$ 
($x_{\rm H}^2+y_{\rm H}^2 \leq a^2$) and $z_{\rm H}=0$~\cite{Cook:1997qc}.

The inverse transformation is obtained as
\begin{align}
r_{\rm BL} =& 
\frac{1}{\sqrt{2}} [r_{\rm H}^2-a^2+((r_{\rm H}^2-a^2)^2+4\,a^2\,z_{\rm H}^2)^{1/2}]^{1/2}
+ M 
\cr
=& R_{\rm H} + M \,,
\cr
t_{\rm BL} =& t_{\rm H} - \frac{(M+\sqrt{M^2-a^2})^2+a^2}{2\,\sqrt{M^2-a^2}} 
\cr & \times
\ln \left|\frac{R_{\rm H}-\sqrt{M^2-a^2}}{R_{\rm H}+\sqrt{M^2-a^2}}\right| \,,
\cr 
\theta_{\rm BL} =& 
\arccos \frac{z_{\rm H}}{r_{\rm BL} - M} = \arccos \frac{z_{\rm H}}{R_{\rm H}}
\,,
\cr 
\phi_{\rm BL} =& \phi_{\rm H} - \phi_{\rm a}
- \frac{a}{r_{+}-r_{-}} 
\,\ln \left|\frac{r_{\rm BL}-r_{+}}{r_{\rm BL}-r_{-}}\right| 
\cr
=& \arctan \frac{y_{\rm H}}{x_{\rm H}} - \arctan \frac{a}{R_{\rm H}}
\cr &
- \frac{a}{2\,\sqrt{M^2-a^2}} 
\,\ln \left|\frac{R_{\rm H}
-\sqrt{M^2-a^2}}{R_{\rm H}+\sqrt{M^2-a^2}}\right| 
\,.
\end{align}
Note here that $r_{\rm H}$ and $R_{\rm H}$ are different.
The Taylor expansion with respect to small $a$ and $M$
of the above relation gives the same equations as in Ref.~\cite{Hergt:2007ha}.
But in Ref.~\cite{Hergt:2007ha}, we find the time coordinate transformation
as $t_{\rm H} = t_{\rm BL}$ because the harmonic coordinates are not unique.
The other transformations remain unchanged from the above equations.

For the evaluation of the ISCO, we turn to Ref.~\cite{Bardeen:1972APJ}
and have the last stable circular orbit
(sometimes referred to as the marginally stable orbit) at
\begin{align}
r_{\rm ms,BL} = & M \{ 3 + Z_2 \mp [(3 - Z_1)(3 + Z_1 + 2 Z_2)]^{1/2}\} \,;
\cr
Z_1 \equiv & 1 + ( 1 - a^2/M^2)^{1/3}
\cr &
\times [(1+ a/M)^{1/3} + (1- a/M)^{1/3}] \,,
\cr
Z_2 \equiv & (3 a^2/M^2 + Z_1^2)^{1/2} \,,
\end{align}
for the BL radial coordinate.
Plugging this radius into the transformation for the CS-H coordinates,
we obtain Table~\ref{tab:ISCO}.

\begin{table}[htb]
\caption{Radius of the marginally stable orbit for various spins in harmonic coordinates.}
\label{tab:ISCO}
\begin{center}
\begin{tabular} {c|c}
\hline\hline
a/M & $r_{\rm ms,H}/M$ \\ \hline
0.9 & 1.59836 \\
0.6 & 2.89200 \\
0.3 & 3.98982 \\
0.0 & 5.00000 \\ \hline\hline
\end{tabular}
\end{center}
\end{table}

%%%%%%%%%%%%%%%%%%%%%%%%%%%%%%%%%%%%%%%%
\section{Computationally effective IZ metric}\label{app:IZtreatment}
%%%%%%%%%%%%%%%%%%%%%%%%%%%%%%%%%%%%%%%%

The metric perturbation in the IZ metric
is described under the ingoing radiation gauge,
$h_{\mu\nu}^{\rm IZ} \ell^\nu = 0$ and $h_{\mu}^{{\rm IZ} \mu} = 0$.
Here, $\ell^\nu$ is the Kinnersley null tetrad~\cite{Yunes:2005ve}.
We can use these five gauge conditions to reduce the computational cost
for the calculation of the IZ metric.
It is noted that all conditions are not independent
and the existence of the gauge condition has been discussed in Ref.~\cite{Price:2006ke}.

In practice, when we calculate $h_{22}^{\rm IZ}$, $h_{23}^{\rm IZ}$,
$h_{24}^{\rm IZ}$, $h_{33}^{\rm IZ}$
and $h_{34}^{\rm IZ}$, the other metric perturbations are derived as
\begin{widetext}
\begin{align}
{h_{11}^{\rm IZ}}=&
\frac{1}{4}\,{\frac { \left( {r}^{2} -{a}^{2}+2\,{a}^{2} \cos^{2} \theta \right)  
\left( {r}^{2}-2\,Mr+{a}^{2} \right) ^{2}{h_{22}^{\rm IZ}}}{ \left( {r}^{2}+{a}^{2} \right)  
\left( {r}^{2}+{a}^{2} \cos^{2} \theta \right) ^{2}}}
+{\frac {a \left( {r}^{2}-2\,Mr+{a}^{2} \right) {h_{24}^{\rm IZ}}}{ \left( {r}^{2}+{a}^{2} \right)  
\left( {r}^{2}+{a}^{2} \cos^{2} \theta \right) }}
-{\frac {{a}^{2} \sin^{2} \theta {h_{33}^{\rm IZ}}}
{ \left( {r}^{2}+{a}^{2} \cos^{2} \theta \right) ^{2}}} \,,
\cr
{h_{12}^{\rm IZ}}=&
-\frac{1}{2}\,{\frac { \left( {r}^{2}-2\,Mr+{a}^{2} \right) {h_{22}^{\rm IZ}}}{{r}^{2}+{a}^{2}}}
-{\frac {a{h_{24}^{\rm IZ}}}{{r}^{2}+{a}^{2}}} \,,
\quad
{h_{13}^{\rm IZ}}=
-\frac{1}{2}\,{\frac { \left( {r}^{2}-2\,Mr+{a}^{2} \right) {h_{23}^{\rm IZ}}}{{r}^{2}+{a}^{2}}}
-{\frac {a{h_{34}^{\rm IZ}}}{{r}^{2}+{a}^{2}}} \,,
\cr
{h_{14}^{\rm IZ}}=& - \frac{1}{4}\,{\frac { \sin^{4} \theta  {a}^{3} 
\left( {r}^{2} + {a}^{2} - 2\,Mr \right) ^{2}{h_{22}^{\rm IZ}}}{ \left( {r}^{2}+{a}^{2} \right)  
\left( {r}^{2}+{a}^{2} \cos^{2} \theta \right) ^{2}}}
-\frac{1}{2}\,{\frac { \left( {r}^{2} -{a}^{2} \cos^{2} \theta +2\,{a}^{2} \right)  
\left( {r}^{2} + {a}^{2} - 2\,Mr \right) {h_{24}^{\rm IZ}}}
{ \left( {r}^{2}+{a}^{2} \cos^{2} \theta \right)  \left( {r}^{2}+{a}^{2} \right) }}
+ {\frac {a \sin^{2} \theta
\left( {r}^{2}+{a}^{2} \right) {h_{33}^{\rm IZ}}}{ \left( {r}^{2}+{a}^{2} \cos^{2} \theta \right) ^{2}}}
\,,
\cr
{h_{44}^{\rm IZ}}=&
{\frac {a \sin^{2} \theta \left( {r}^{2}-2\,Mr+{a}^{2} \right) {h_{24}^{\rm IZ}}}
{{r}^{2}+{a}^{2} \cos^{2} \theta  }}
-\frac{1}{4}\,{\frac {{a}^{2} \sin^{4} \theta
\left( {r}^{2}-2\,Mr+{a}^{2} \right) ^{2} {h_{22}^{\rm IZ}}}
{ \left( {r}^{2}+{a}^{2} \cos^{2} \theta \right) ^{2}}}
-{\frac { \sin^{2} \theta \left( {r}^{2}+{a}^{2} \right) ^{2}{h_{33}^{\rm IZ}}}
{ \left( {r}^{2}+{a}^{2} \cos^{2} \theta \right) ^{2}}} \,.
\end{align}
\end{widetext}

%%%%%%%%%%%%%%%%%%%%%%%%%%%%%%%%%%%%%%%%

\bibliographystyle{apsrev4-1}
\bibliography{./references,references}

\end{document}